\documentclass[]{aa}
\usepackage[utf8]{inputenc}
\usepackage[T1]{fontenc}
\usepackage{txfonts}
\usepackage{graphicx}
\usepackage{float} 
\usepackage{amsmath}
\usepackage{amssymb}
\usepackage[colorlinks=true, linkcolor=blue, filecolor=blue, citecolor=blue, urlcolor=blue]{hyperref}    

\usepackage[dvipsnames]{xcolor}

\usepackage{placeins}

\usepackage{siunitx}
\usepackage[normalem]{ulem}

\usepackage{cancel}
\usepackage{booktabs}
\usepackage{multirow}
\usepackage{adjustbox}

\makeatletter
\renewcommand*\aa@pageof{, page \thepage{} of \pageref*{LastPage}}
\makeatother

\begin{document}
\title{The role of mass transfer efficiency in stability criteria: Implementation in \texttt{SEVN} and a test on blue stragglers and binary compact objects}
\titlerunning{MT stability in \texttt{SEVN}}
\author{ 
        M. Echeveste\inst{\ref{oaa}},
        G. J. Escobar\inst{\ref{aca}, \ref{ula}},
        G. Iorio\inst{ \ref{daul}},
        E.~Pancino\inst{\ref{oaa}}, 
        M. Mapelli\inst{\ref{zah}, \ref{izwr}, \ref{gg}, \ref{ifnf}},
        D.~Alvarez Garay\inst{\ref{oaa}},
        A.~Avdeeva\inst{\ref{oaa}},
        E.~Leitinger\inst{\ref{oaa}, \ref{unibo}},
        S. Nedhath\inst{\ref{oaa}, \ref{unifi}},
        S. Rani\inst{\ref{oaa}},
        E. Reggiani\inst{\ref{oaa}, \ref{unifi}},
        N.~Sanna\inst{\ref{oaa}},
        S. Saracino\inst{\ref{oaa}},
        L. Steinbauer\inst{\ref{oaa}, \ref{unifi}},
        A.~Turchi\inst{\ref{oaa}}
        }
\authorrunning{M.~Echeveste et al.}

\institute{INAF -- Osservatorio Astrofisico di Arcetri, Largo E. Fermi 5, 50125 Firenze, Italy\label{oaa}
\and Instituto de Astrof\'isica de Canarias, E-38205 La Laguna, Tenerife, Spain\label{aca}
\and Universidad de La Laguna, Dpto. Astrofísica, E-38206 La Laguna, Tenerife, Spain\label{ula}
\and Institut de Ciències del Cosmos (ICCUB), Universitat de Barcelona (UB), c. Martí i Franquès, 1, 08028 Barcelona, Spain\label{daul}
\and Universit\"at Heidelberg, Zentrum f\"ur Astronomie (ZAH), Institut f\"ur Theoretische Astrophysik, Albert Ueberle Str. 2, 69120, Heidelberg, Germany\label{zah}
\and Universit\"at Heidelberg, Interdiszipli\"ares Zentrum f\"ur Wissenschaftliches Rechnen, D-69120 Heidelberg, Germany\label{izwr}
\and Dipartimento di Fisica e Astronomia Galileo Galilei, Università di Padova, Vicolo dell’Osservatorio 3, I–35122 Padova, Italy\label{gg}
\and INFN, Sezione di Padova, Via Marzolo 8, I--35131 Padova, Italy\label{ifnf}
\and Dipartimento di Fisica e Astronomia, Universit{\`a} degli Studi di Bologna, Via Gobetti 93/2, 40129 Bologna, Italy\label{unibo}
\and Dipartimento di Fisica e Astronomia, Universit{\`a} di Firenze,  Via G. Sansone 1, 50019 Sesto Fiorentino, FI, Italy\label{unifi}
}

\date{Received: \today}

\abstract
{The stability of mass transfer through Roche-lobe overflow plays a key role in shaping the outcome of binary interactions. However, the criterion for mass transfer stability remains one of the main open questions in the theory of binary evolution.}
{Our aim was to close this knowledge gap by developing a mass transfer stability criterion that accounts for mass and angular momentum loss, and implemented it in the population synthesis code \texttt{SEVN}. We assessed its impact relative to the standard formalism used in \texttt{SEVN}, using blue straggler stars and binary compact objects as illustrative cases.}
{We derived an expression for the response of the Roche-lobe radius to mass loss in the general case where the mass and angular momentum of the system are not conserved. On the basis of this formulation, we constructed a new mass transfer stability criterion that modifies the standard approach only through the Roche-lobe response term. We performed population synthesis simulations with the \texttt{SEVN} code to explore the effects of the new stability criterion.}
{The new criterion allows stable mass transfer in binaries with higher donor-to-accretor mass ratios, leading to an overall increase in the predicted number of blue stragglers and promoting their formation in wider orbits. This contributes to reconciling the differences between theory and observations. For binary compact objects, the impact of the new stability prescription varies across system types, with the strongest effects occurring in binaries containing at least one neutron star. In particular, for low mass transfer efficiency, the new prescription enhances the contribution of channels involving stable mass transfer and leads to a larger number of systems, including gravitational wave progenitors.}
{The inclusion of a new, simple, yet more consistent prescription for mass transfer stability has proven that refining this criterion can significantly improve our understanding of the formation channels of specific stellar populations.}


\maketitle{}

\section{Introduction}

The wide range of physical processes occurring in interacting binary systems gives rise to multiple evolutionary pathways and a rich variety of stellar outcomes and observable phenomena. These interactions play a fundamental role in shaping the properties and evolution of stellar populations; therefore, studying binary interactions is essential to understanding the formation and evolution of these populations.

Among these interactions, one of the most important is mass transfer (MT) through Roche-lobe overflow. The stability of this process plays a pivotal role in determining the outcome of binary evolution (see, e.g., \citealt{Ge2024arXiv241117333G}). If MT is unstable, the system is thought to go through a common envelope (CE) phase in which the loss of orbital energy leads to the shrinking of the orbit and to the unbinding of the envelope. If the process is efficient enough, the binary expels the CE and forms a tight binary \citep{Paczynski1976IAUS...73...75P}; otherwise, there is a  merger (see, e.g., \citealt{Ivanova2013A&ARv..21...59I}). On the other hand, stable MT leads to detached configurations, and a fraction of the mass can be lost from the system in such a process, removing angular momentum. How much mass is successfully accreted by the accretor and how much is lost from the system is under debate, and its determination constitutes one of the major observational and theoretical challenges in the study of interacting binaries \citep{DeMink2007,Eldridge2008,Shao2014}. 

The stability of MT through Roche-lobe overflow depends on how the radius of the donor star evolves relative to its Roche-lobe radius (see, e.g., Equations 2 to 11 in \citealt{ge2010ApJ...717..724G}). For MT to remain stable, the donor’s radius must adjust to mass loss without leading to a runaway increase of the Roche-lobe overflow. While the response of the star's radius to mass loss depends on the stellar structure, how its Roche-lobe radius reacts depends on the binary's configuration and whether the MT process occurs while conserving the total mass and angular momentum of the system (e.g., \citealt{soberman1997}). Classical prescriptions used in population synthesis codes define the transition between stable and unstable MT by assuming that the Roche-lobe radius of the donor evolves under conservative MT \citep{Tout1997, Hurley2002}. In this work, we adopt the formalism derived in \citet{Echeveste2026} and \citet{Escobar2026}, where an expression for the response of the Roche-lobe radius to mass loss in the general case of non-conservative MT was presented. An analogous approach was adopted by \citet{Willcox2023ApJ...958..138W} and implemented in the binary population synthesis code \texttt{COMPAS} \citep{Stevenson2017NatCo...814906S,VignaGomez2018MNRAS.481.4009V}, a code built to simulate massive stellar binaries. They found that the fraction of systems that undergo unstable MT and mergers depends on the stability criteria and is governed by the angular momentum loss (AML) mode. Building on this framework, we have developed a new prescription to compute the stability criterion for Roche-lobe overflow MT and implemented it in the population synthesis code \texttt{SEVN}\footnote{https://sevncodes.gitlab.io/sevn/, version 2.16.0}. We investigate how our new formulation affects the stability of MT by exploring different values of the MT efficiency and two distinct modes of AML (either from the vicinity of the donor or from the accretor star). We generate different sets of large populations and use blue straggler stars (BSSs) and binary compact objects (BCOs) as test cases to evaluate the effects of the new prescription.

Population synthesis simulations provide a powerful tool for investigating the properties of large stellar populations at any time of their evolution and comparing them with observations. This tool has been continuously growing during the past 20 years and has been applied in several research lines, such as in the study of progenitors of gravitational wave (GW) sources \citep{Banerjee2010MNRAS.402..371B,rodriguez2016PhRvD..93h4029R,Chatterjee2017ApJ...834...68C,Neijssel2019MNRAS.490.3740N,kremer2020ApJS..247...48K,Riley2021,vanSon2022ApJ...931...17V,Costa2023,picco2024A&A...681A..31P,Mestichelli2024}, the effect of cluster dynamics on the formation of binaries containing black holes \citep{2016MNRAS.462.3302E,belczynski2020A&A...636A.104B,Olejak2021,bavera2021A&A...647A.153B,Chattopadhyay2021MNRAS.504.3682C,Broekgaarden2021MNRAS.508.5028B,Torniamenti2022, Rastello2023, MarinPina2024}, double white dwarfs \citep{Nelemans2001A&A...365..491N,Toonen2012A&A...546A..70T,Breivik2018ApJ...854L...1B,korol2020A&A...638A.153K, korol2022MNRAS.515.1228K,li2023A&A...669A..82L}, the spectral modeling of galaxies \citep{Eldridge2017PASA...34...58E,Lecroq2024MNRAS.527.9480L}, and the formation of blue straggler stars (e.g., \citealt{2005MNRAS.363..293H,Geller_2012,Leiner_2021}).

Most population synthesis codes for binary evolution are built on the basis of the prescriptions for stellar evolution and binary processes by \citet{Hurley2000, Hurley2002}, which are based on polynomial fits to stellar evolution tracks computed by \citet{Pols1998}. This method set up a paradigm in which each code has been subsequently tailored according to each specific objective (e.g., \texttt{BSE}, \citealt{Hurley2000, Hurley2002}; \texttt{MOBSE}, \citealt{Mapelli2017,Giacobbo2018}; \texttt{STARTRACK},  \citealt{Belczynski2002ApJ...572..407B,Belczynski2008ApJS..174..223B}; \texttt{SEBA}, \citealt{Portegies1996A&A...309..179P,Toonen2012A&A...546A..70T}; \texttt{COMPAS}, \citealt{Riley2022,Rodriguez-Segovia2025_I,Rodriguez-Segovia2025_II}; \texttt{COSMIC}, \citealt{Breivik2020ApJ...898...71B}; \texttt{BINARY\_C}, \citealt{Izzard2004,izzard2006A&A...460..565I,Izzard2009,Izzard2018}). In recent years, a new paradigm has been developed, based on the interpolation of modern stellar evolutionary tracks on the fly. This approach enables the evolution of stars to be computed with higher precision than traditional polynomial fitting, while keeping the computational cost at a reasonable level (e.g., \texttt{Brussels code}, \citealt{dedodner2004NewAR..48..861D}, \texttt{COMBINE}, \citealt{Kruckow2018}; \texttt{METISSE}, \citealt{Agrawal2020, Agrawal2023}; \texttt{SEVN}, \citealt{spera2017MNRAS.470.4739S}, \citealt{Spera2019}, \citealt{Mapelli2020}, \citealt{Iorio2023}; and \texttt{POSYDON}, \citealt{Fragos2023}, \citealt{Andrews2024arXiv} and \texttt{BPASS}, \citealt{Eldridge2017PASA...34...58E}, both of which also account for binary evolution by interpolating detailed models). In particular, this new paradigm allows one to easily explore and compare the predictions from different stellar models by simply choosing the stellar tracks that come from the corresponding detailed simulations.

\texttt{SEVN} is a public population synthesis code that interpolates stellar evolutionary tracks on the fly. The current public release includes evolutionary tables computed with \texttt{MESA} \citep{Paxton2011,Paxton2013,Paxton2015}, specifically the \texttt{MIST} tracks \citep{choi2016ApJ...823..102C}, as well as \texttt{PARSEC} tracks \citep{Bressan_2012,Chen2015} from the \texttt{V2} dataset \citep{costa2025A&A...694A.193C}. In addition, \texttt{SEVN} includes updated prescriptions for computing the outcome and the effect of several binary processes, such as tides, supernova kicks, rotation, mass transfer, and common-envelope evolution, among others. A complete description of \texttt{SEVN} can be found in \citet{Iorio2023}.

This paper is organized as follows. In Sect.~\ref{sec:2} we present the new stability criterion for MT through Roche-lobe overflow. Sect.~\ref{sec:results} describes the impact of this formulation on the population of BSSs and binary compact objects generated with the code \texttt{SEVN}. Finally, in Sect.~\ref{sec:conclusion} we elaborate on our conclusions.

\section{Derivation of the mass transfer stability criterion in the non-conservative case}
\label{sec:2}

The stability of MT is primarily determined by the response of the donor star to mass loss and by how this response compares with the variation of its Roche-lobe radius. For MT to remain stable, the donor must avoid an ever-increasing overflow of its Roche lobe; thus, its radius should not expand more rapidly (or contract more slowly) than the Roche-lobe radius. 

The star's response to mass loss on a dynamical timescale can be quantified through the adiabatic mass–radius exponent, i.e., the change of radius of the donor needed to reach a new hydrostatic equilibrium as a consequence of mass loss,

\begin{equation}
    \zeta_{\rm ad} = \frac{d \ln R_{\rm d}}{ d \ln M_{\rm d}} \bigg |_{\rm ad},
\end{equation}

\noindent where $M_{\rm d}$ and $R_{\rm d}$ denote the donor’s mass and radius, respectively (e.g., \citealt{webbink1985ibs..book...39W}). This exponent reflects the internal structure of the donor and, in particular, the dominant energy transport mechanism in the envelope: Convective envelopes generally expand in response to mass loss, whereas radiative envelopes contract. 

The evolution of the radius of the Roche lobe with mass loss, $R_{\rm L,d}$, is described by

\begin{equation}
    \zeta_{\rm L} = \frac{d \ln R_{\rm L,d}}{ d \ln M_{\rm d}},
\end{equation}

\noindent which depends on the geometry of the binary (e.g., \citealt{webbink1985ibs..book...39W}). A comparison of the two indices provides the criterion for the stability of Roche-lobe overflow: MT is dynamically unstable when $\zeta_{\rm L} > \zeta_{\rm ad}$ and stable otherwise (see, e.g., \citealt{soberman1997,ge2010ApJ...717..724G}).

Most population synthesis codes do not calculate $\zeta_{\rm L}$ and $\zeta_{\rm ad}$ on the fly but rather approximate this criterion with the definition of a critical mass ration $q_{\rm c}$, above which MT is assumed to be unstable. The main reason is that the Roche-lobe response is controlled by the mass ratio $q$, and therefore the stability criterion can be reformulated as a threshold in $q$.

The critical mass ratio $q_{\rm c}$ that represents the limit between stable and unstable MT used in \texttt{SEVN} is given in \citet{Hurley2002}. This $q_{\rm c}$ takes different values according to the phase of the donor star and is defined by $\zeta_{\rm ad}=\zeta_{\rm L}$, where $\zeta_{\rm L}=\zeta_{\rm L}(q)=2.13 q - 1.67$ \citep{Tout1997}, with $q = M_{\rm d} / M_{\rm a}$ and $M_{\rm a}$ the mass of the accretor. This expression for $\zeta_{\rm L}$ is deduced under conservative assumptions, i.e., no mass nor angular momentum is lost during the MT process, and therefore it depends only on $q$.  

Here, we derive a new expression for $\zeta_{\rm L}$ that relaxes the conservative assumption. To do so, we used the same approximation for the Roche lobe used in Tout's work and, therefore, in Hurley's calculation of $q_{\rm c}$. It is given by \citet{Eggleton1983} and is valid for $0.1 < q < 10$:

\begin{equation}
    \frac{R_{\rm L,d}}{a} = 0.44 \frac{q^{0.33}}{(1+q)^{0.2}},
    \label{eq:rl}
\end{equation}

\noindent where $a$ is the semi-major axis. This approximation assumes the Roche lobe to be a perfectly spherical surface.

Assuming the binary is circularized, the orbital evolution of the system follows

\begin{equation}
    2 \frac{\dot{J}}{J} = \frac{\dot{a}}{a} + 2\frac{\dot{M_{\rm d}}}{M_{\rm d}} + 2\frac{\dot{M_{\rm a}}}{M_{\rm a}} - \frac{\dot{M_{\rm d}}+\dot{M_{\rm a}}}{M_{\rm d}+M_{\rm a}},
    \label{eq:dJ}
\end{equation}

\noindent where $\dot{M_{\rm d}}$ and $\dot{M_{\rm a}}$ are the mass-loss rate of the donor and the accretion rate of the accretor, respectively, and $J$ is the total angular momentum. To describe non-conservative MT, we introduce the parameters $\beta$ and $\gamma$ (see, e.g., \citealt{Pols2018}) as

\begin{equation} \label{eq:bg}
\begin{gathered}
    \dot{M}_{\rm a} = - \beta \dot{M}_{\rm d}, \\
    h_{\rm loss} = \frac{\dot{J}}{ \dot{M}_{\rm a} + \dot{M}_{\rm d}} = \gamma \frac{J}{M_{\rm a} + M_{\rm d}}.
\end{gathered}
\end{equation}

\noindent The parameter $\beta$ measures the efficiency of the MT, with $\beta=1$ being the conservative case. The matter that is not accreted is lost from the system carrying away angular momentum. The quantity $h_{\rm loss}$ represents the specific angular momentum of the ejected matter, defined as $\gamma$ times the specific angular momentum of the binary\footnote{Notice that the standard formalism in \texttt{SEVN} uses the reduced mass in the definition of $\gamma$ (see appendix \textit{A4.2} in \citealt{Iorio2023}), same as in \citet{2023pbse.book.....T}.}. Using these definitions, equation \ref{eq:dJ} can be written as

\begin{equation}
    \frac{\dot{a}}{a}= -2 \frac{\dot{M_{\rm d}}}{M_{\rm d}} \bigg [1- \beta \frac{M_{\rm d}}{M_{\rm a}} - (1-\beta)(\gamma + 1/2) \frac{M_{\rm d}}{M_{\rm d} + M_{\rm a}} \bigg].
    \label{eq:abg}
\end{equation}

Using Equations \ref{eq:rl} and \ref{eq:abg}, we deduced the expression

\begin{equation}
\begin{split}
    \zeta_{\rm L} (q,\beta,\gamma)=& -1.67 + 2(1-\beta)(\gamma + 1/2) \frac{q}{q+1} \\
    & + q \beta \left[ 2.33 - \frac{0.2(1+q\beta)}{\beta(1+q)}\right],
\end{split}
\label{eq:rlzet}
\end{equation}

\noindent which for $\beta = 1$ reduces to that of \citet{Tout1997}. 

If the angular momentum is lost isotropically from the vicinity of the accretor (usually referred to as isotropic re-emission), then the specific angular momentum of the ejected matter is $h_{\rm loss} = a_{\rm a}^2 \omega$, where $a_{\rm a} = a M_{\rm d}/(M_{\rm d}+M_{\rm a})$ is the relative orbit of the accretor around the center of mass, and $\omega$ is its frequency. In this case,

\begin{equation}
    h_{\rm loss} =  a_{\rm a}^2 \omega = \bigg (\frac{M_{\rm d}}{M_{\rm d} + M_{\rm a}} \bigg)^2 
    \sqrt{G (M_{\rm d} + M_{\rm a})a}.
\end{equation}

\noindent It can be shown that this case corresponds to $\gamma = q$. Analogously, if the mass is lost from the vicinity of the donor star (fast mode or Jeans mode), it can be verified that $\gamma = q^{-1}$. In this work, we will consider these two AML modes.

Having introduced the relevant equations, we can now build a criterion for the stability of the MT. To make a direct comparison with Hurley's $q_{\rm c}$, we used $\zeta_{\rm ad}$: 

\begin{equation}
    \zeta_{\rm ad}=
    \begin{cases}
    -0.1896, & \text{low mass MS,} \\%(k=0)} \\
    4.72, & \text{MS, CHeB, HeMS,} \\%(k=1,4,7)} \\
    6.85, & \text{HG,} \\ %(k=2)}
    \zeta_{\rm ad, GB}, & \text{GB, EAGB,} \\% (k=3,5)} \\
    0, & \text{HeHG,} \\% (k=8)} \\
    -1/3, & \text{WD,} \\% (k=10,11,12)} 
    \end{cases} 
    \label{eq:adzet}
\end{equation}

\noindent where MS is main sequence, HG is Hertzsprung-gap, GB is giant branch, CHeB is core-helium burning, EAGB is early asymptotic giant branch, HeMS is naked-helium MS, HeHG is naked-helium HG, and WD is white dwarf, as defined in \citet{Hurley2000}. For GB stars, we used the formula

\begin{equation}
    \zeta_{\rm ad, GB} =  -0.89894 + 0.71/(1-m_{\rm c}),
\label{eq:adzetGB}
\end{equation}

\noindent with $m_{\rm c}$ the fraction of mass in the core. This expression is deduced from the equation for $q_{\rm c}$ given by \citet{Webbink1988}, \citet{Hjellming1987}. 

For each donor evolutionary phase, the critical mass ratio $q_{\rm c}$ reported by \citet{Hurley2002} and \citet{Iorio2023} is derived by equating the values of $\zeta_{\rm ad}$ in Equation \ref{eq:adzet} and \ref{eq:adzetGB} with the analytic expression of $\zeta_{\rm L}(q)$ provided by \citet{Tout1997}. In previous versions of \texttt{SEVN}, these $q_{\rm c}$ values are used to distinguish between stable and unstable MT. In what follows, we refer to this implementation as the HW criterion, named after Hurley and Webbink. Here, we design a different method to make the distinction between stable and unstable MT: We directly compare the value of $\zeta_{\rm L} (q,\beta,\gamma)$ in Equation \ref{eq:rlzet} with the values of $\zeta_{\rm ad}$ in Equations \ref{eq:adzet} and \ref{eq:adzetGB}. Then, in every time step, the code performs this comparison and distinguishes between stable, i.e., $\zeta_{\rm L} (q,\beta,\gamma) \leq \zeta_{\rm ad}$, and unstable MT, i.e., $\zeta_{\rm L} (q,\beta,\gamma) > \zeta_{\rm ad}$. In what follows, we refer to the stability criterion given by $\zeta_{\rm ad}=\zeta_{\rm L}(q,\beta,\gamma)$ as the BD criterion, named after the idea of a $\beta$-dependent treatment. Notice that the difference between Hurley's prescription and the one we deduce here relies on the expression for $\zeta_{\rm L}$ only, since the adopted values of $\zeta_{\rm ad}$ are implicit in the Hurley's $q_{\rm c}$ values. In particular, for the conservative case, the new implementation is equivalent to the HW criterion, since Equation \ref{eq:rlzet} reduces to that provided by \citet{Tout1997} when $\beta = 1$. In the following sections, we will compare the BD with the standard HW criterion. It is also worth pointing out that the new BD implementation for the MT stability is more self-consistent than the standard HW prescription, since it considers the effect of $\beta$ and $\gamma$ also in the MT stability, while the HW criterion assumes conservative MT even when choosing $\beta \ne 1$ as an input parameter. In addition, the BD implementation can easily take any other value of $\zeta_{\rm ad}$, and therefore it is more flexible to future improvements on the calculation of these coefficients.

\begin{figure}
    \centering
    \includegraphics[width=1.0\linewidth]{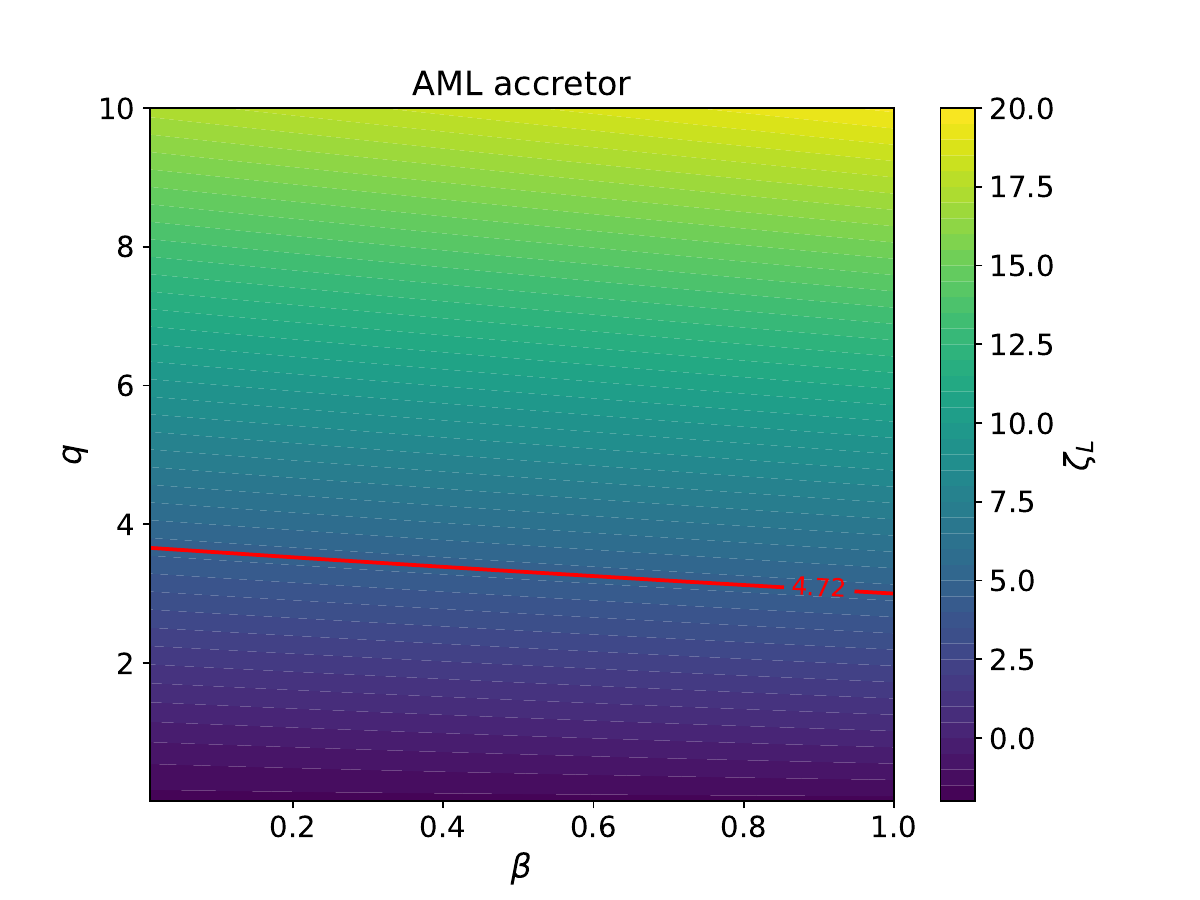}
    \caption{Value of $\zeta_{\rm L}$ as a function of mass ratio, $q$, and MT efficiency, $\beta$, assuming AML from the vicinity of the accretor. A reference level contour at 4.72, corresponding to the $\zeta_{\rm L}$ value for MS donors in the HW formalism, is shown in red. }
    \label{fig:ZetaGB_a}
\end{figure}

\begin{figure}
    \centering
    \includegraphics[width=1.0\linewidth]{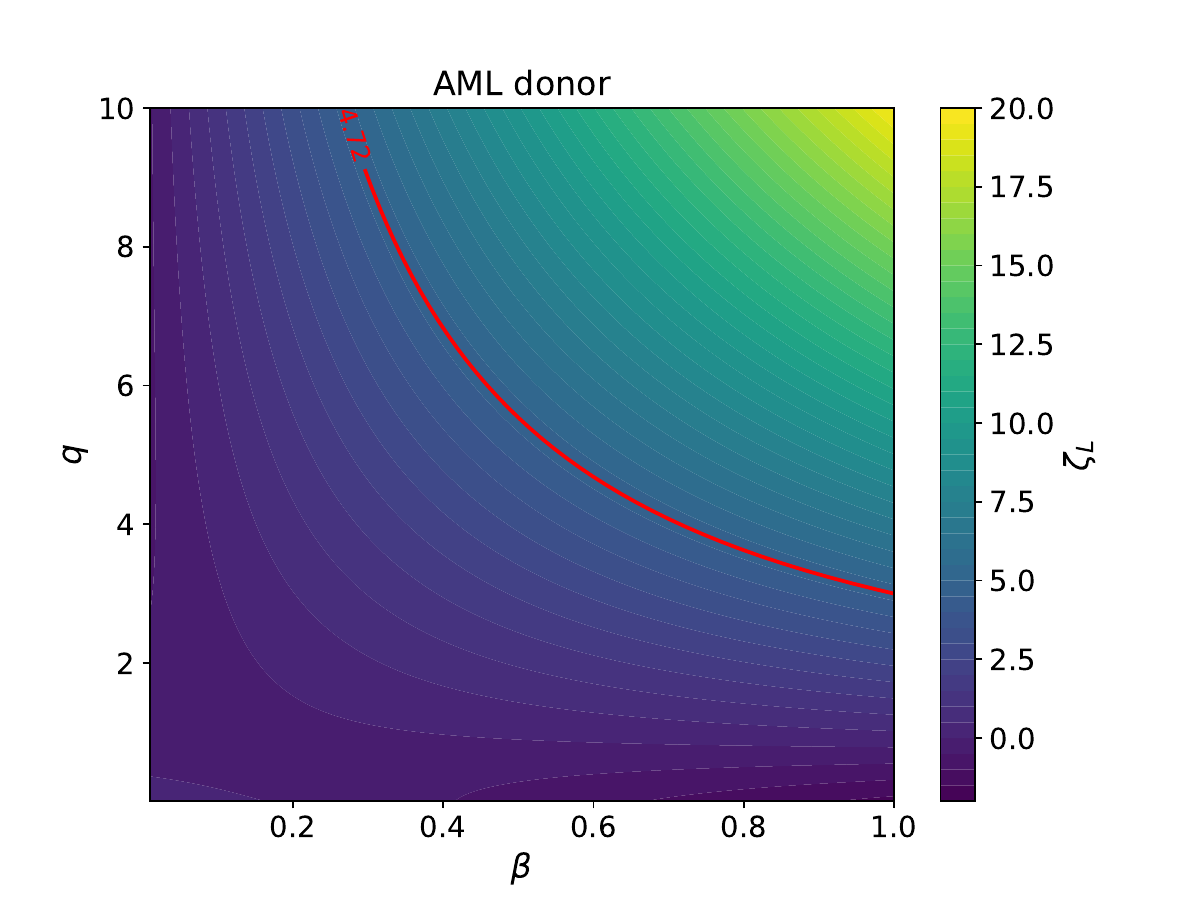}
    \caption{Same as Figure~\ref{fig:ZetaGB_a} but for AML from the vicinity of the donor. }
    \label{fig:ZetaGB_d}
\end{figure}

Figures~\ref{fig:ZetaGB_a} and \ref{fig:ZetaGB_d} show the values of $\zeta_{\rm L} (q,\beta,\gamma)$ as a function of $q$ and $\beta$ and fixed AML mode, i.e., from the vicinity of the accretor and donor. The contour levels trace constant values of $\zeta_{\rm L}$, allowing us to quantify how the critical mass ratio $q_{\rm c}$ varies with $\beta$ and to compare the HW and BD stability criteria. Recall that MT is stable when $\zeta_{\rm L} (q,\beta,\gamma) \leq \zeta_{\rm ad}$ and unstable otherwise. As an example, for MS donors with $\zeta_{\rm ad} = 4.72$, the critical mass ratio $q_{\rm c}$ increases from 3 for $\beta = 1$ (corresponding to the HW criterion) to $\sim 3.6$ for $\beta = 0.1$ if the AML is from the accretor. If the AML is from the donor, $q_ {\rm c}$ increases to $\simeq 10$ when $\beta = 0.26$ and MT remains stable for lower values of $\beta$. This behavior reflects the dependence of the Roche-lobe response on the angular-momentum loss mechanism, as a significant part of the evolution of the Roche lobe depends on how fast the semi-major axis shrinks as a response to the angular momentum leaving the system to tether with the non accreted mass. Since in most of the parameter space $q>1$, the donor is the more massive component and orbits closer to the binary barycenter, carrying less orbital angular momentum than the accretor. As a result, when mass is lost from the donor, only a small amount of angular momentum is removed from the system and the semi-major axis does not shrink significantly. In contrast, mass loss from the accretor extracts a larger amount of angular momentum, leading to a stronger orbital contraction and, consequently, lower values of $q_{\rm c}$. In general, over a wide range of donor $\zeta_{\rm ad}$ values, the BD stability criterion leads to higher $q_{\rm c}$ values as $\beta$ decreases (with the exception of $\zeta_{\rm ad} \lesssim 0$ in the donor AML case). As a consequence, the HW criterion effectively provides a lower bound on $q_{\rm c}$, and population-synthesis studies based on this formalism may overestimate the occurrence of unstable MT episodes.

\section{Population synthesis with \texttt{SEVN}}
\label{sec:results}

The simulations were carried out using \texttt{SEVN} version 2.16.0, modified to include the Roche-lobe overflow MT stability criterion described in Sect.~\ref{sec:2}. Aside from the MT efficiency and the angular momentum loss prescriptions, both of which are explicitly explored here, all other binary processes and parameters were kept at their default values. Some of these are, e.g., equilibrium tides \citep{Hut1981A&A....99..126H}; circularization conserving the binary angular momentum; Roche-lobe overflow formalism and winds accretion by \citet{Hurley2002}; the $\alpha-\lambda$ energy formalism for the CE evolution \citep{Hurley2002} with $\alpha_{\mathrm{CE}} = 3$ and $\lambda$ taking the same values as in \texttt{BSE} and \texttt{MOBSE} (based on the \citealt{claeys2014A&A...563A..83C} prescriptions, see Appendix 1.4 in \citealt{Iorio2023}); supernova kicks following \citet{Giacobbo2020ApJ...891..141G}; GW decay by \citet{Peters1964}; collision at periastron following \citet{Hurley2002}, which can produce a CE or a direct merger depending on the stellar phases; and the delayed supernova explosion mechanism \citep{Fryer2012ApJ...749...91F} with the correction for pair instability to the remnant mass by \citet{Mapelli2020}.

We have used different stellar evolution tracks to study the formation of BSSs and BCOs. The main reason for this is that the range of masses to explore in each case is different. In the case of BSSs, low-mass stars are necessary to account for the formation of these objects, and the MIST tracks from \citet{choi2016ApJ...823..102C} provide better coverage of such stars. In this case, we use the family of tracks which only includes tracks with metallicities that have almost uniform $M_{\rm ZAMS}$ coverage and also includes tracks that do not reach the end of the AGB. For such incomplete tracks, the envelope is manually stripped, setting the total mass of the star to 1.02 times the mass of the He content. In the case of simulations involving binary compact objects, we employed the set of PARSEC tracks with overshooting parameter $\lambda = 0.5$, which includes stars with masses up to $600~M_{\odot}$. For He stars, we used the default PARSEC tracks.

The following subsections present simulations for two representative test cases, based on BSSs and BCOs. For each case, we first describe the specific setup of the simulations, followed by the main results and their physical implications. This structure allows us to highlight how the new stability prescription affects different binary populations.

\subsection{Blue straggler stars}

Blue straggler stars are stellar objects historically identified by their peculiar location in the color-magnitude diagram, brighter and bluer than the turn-off point \citep{sandage53}, a region that stars evolved in isolation should not populate. It is widely believed that these exotic objects are MS stars that acquired mass after their formation, with binary MT and mergers two of the possible formation channels usually considered \citep{McCrea_1964,Tian_2006,Andronov_2006,Chen2008,2008MNRAS.387.1416C,Knigge_2009}.

In this section, we investigate how the formation of BSSs is influenced by the adoption of the new MT stability criterion (BD, as defined in Sect. \ref{sec:2}). To isolate its impact, we compare two sets of simulations that are identical in all aspects except for the treatment of MT stability, where one employs the BD prescription and the other the HW prescription. The simulations were performed on an initial population of $10^5$ binaries in the zero-age main sequence (ZAMS) and with a metallicity of $Z=0.014$. Orbital periods and eccentricities were generated following a uniform linear and logarithmic distribution, respectively. The masses of both stars in the binary were drawn from a Kroupa initial mass function \citep[IMF,][]{Kroupa2001}, with masses from 0.7 to 10 $M_\odot$, and then randomly paired. The total mass of the simulated binaries is $3.44 \times 10^5 M_{\odot}$, which corresponds to an effective total mass of $3.15 \times 10^6 M_{\odot}$ after correcting for the incomplete IMF sampling\footnote{The correction factor is calculated as the ratio between the total mass of the simulation, which in this case corresponds to an IMF integrated between 0.7 and $10 M_{\odot}$, and the mass of a complete population, which we calculate as the integral of the IMF between 0.08 and $150 M_{\odot}$. We consider the IMF of one star independent from the other and assume a binary fraction of 1.}. 

We performed a total of 36 sets of simulations, i.e., using nine values of $\beta$ (from 0.1 to 0.9 with a step of 0.1), two AML modes (from the accretor and from the donor), and two MT stability criteria (BD and HW). 

\subsubsection{BSS results}

We select BSSs from our simulations considering two possible scenarios: 1) a BSS is the merger of two MS stars and is more massive than the turnoff mass, and 2) a BSS is a MS star in a binary system that has acquired sufficient mass to be more massive than the turnoff mass. We evaluate the formation of BSSs at different ages (from 0.5 to 10 Gyr), selecting the turnoff mass from the isochrones provided by the MIST tracks. 

In what follows, we report the formation efficiency of the BSSs found in both scenarios. The formation efficiency is defined as

\begin{equation}
    \eta_{\rm BSS} = \frac{N_{\rm BSS}}{M_{\rm pop}},
\end{equation}

\noindent with $N_{\rm BSS}$ the number of BSSs in the simulation and $M_{\rm pop}$ the effective total mass.

\begin{figure}
    \centering
    \includegraphics[width=1.0\linewidth]{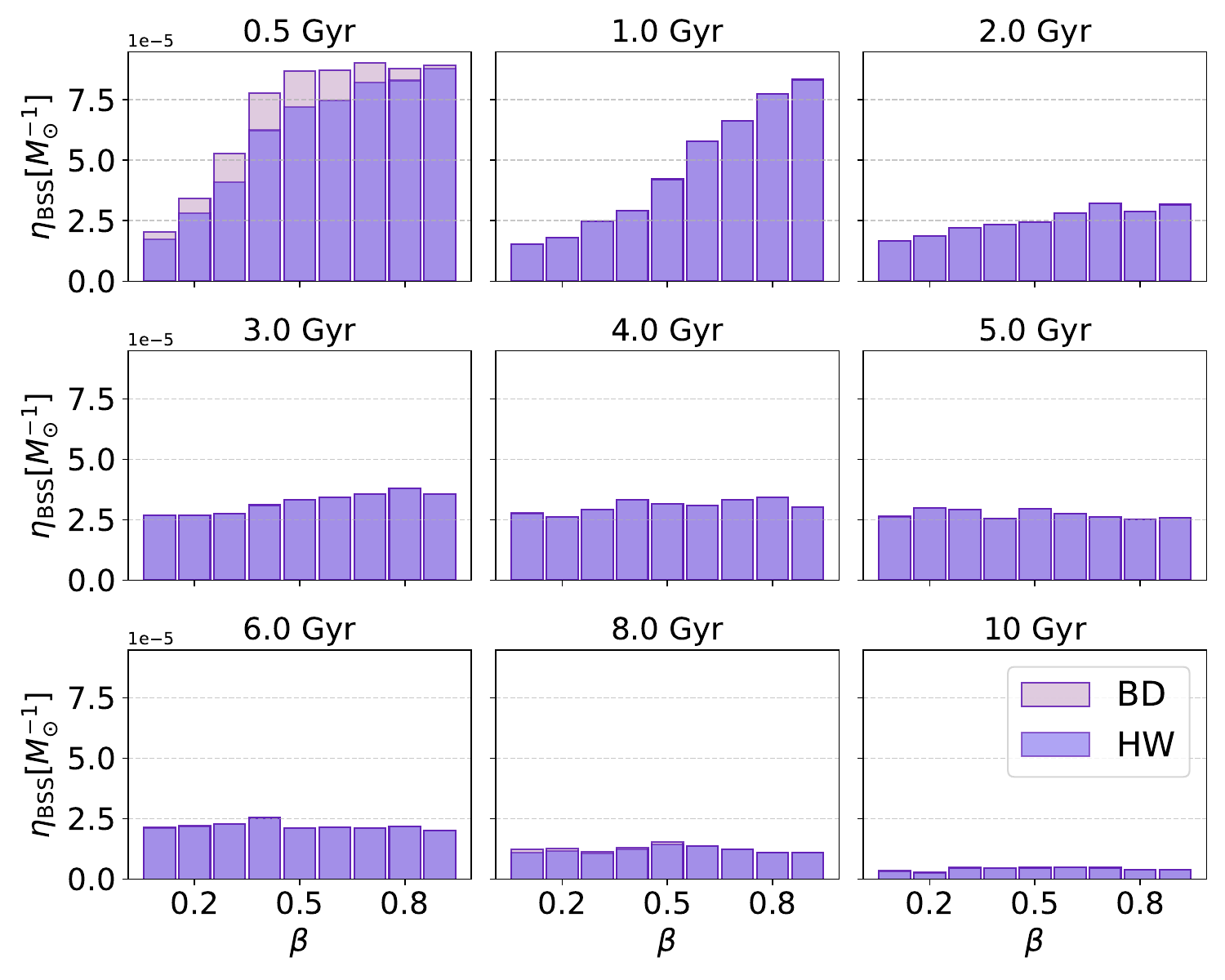}
    \caption{Formation efficiency of BSSs produced through the merger of two MS stars (first scenario) as a function of the MT efficiency $\beta$. Each panel corresponds to a different age. Results obtained using the BD prescription are shown in light purple, while those computed with the HW prescription are shown in dark purple. In all cases, angular momentum is assumed to be lost from the vicinity of the accretor. }
    \label{fig:hist_merge_a}
\end{figure}

\begin{figure}
    \centering
    \includegraphics[width=1.0\linewidth]{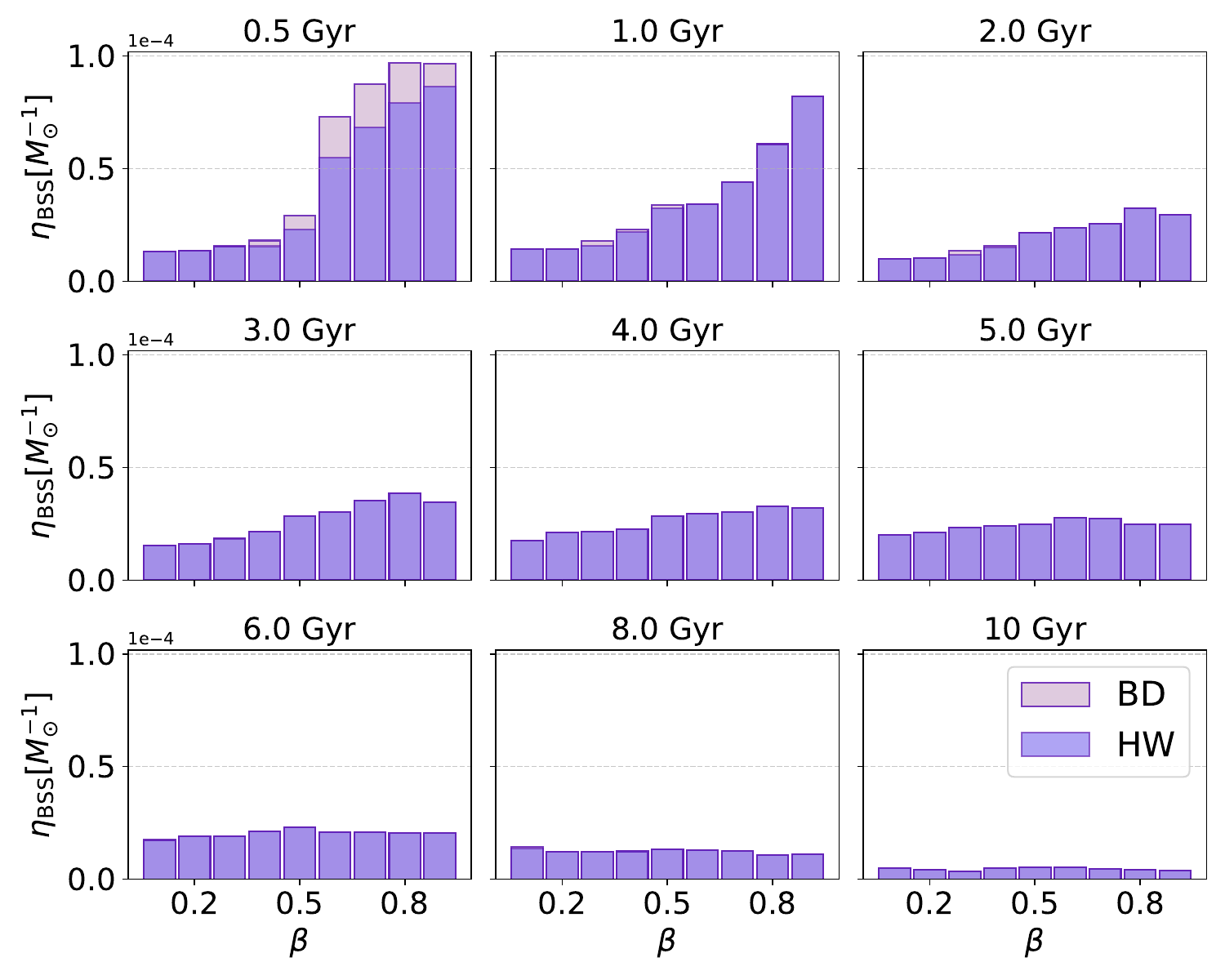}
    \caption{Same as Figure \ref{fig:hist_merge_a} but for the case in which AML is from the vicinity of the donor. }
    \label{fig:hist_merge_d}
\end{figure}

\begin{figure}
    \centering
    \includegraphics[width=1.0\linewidth]{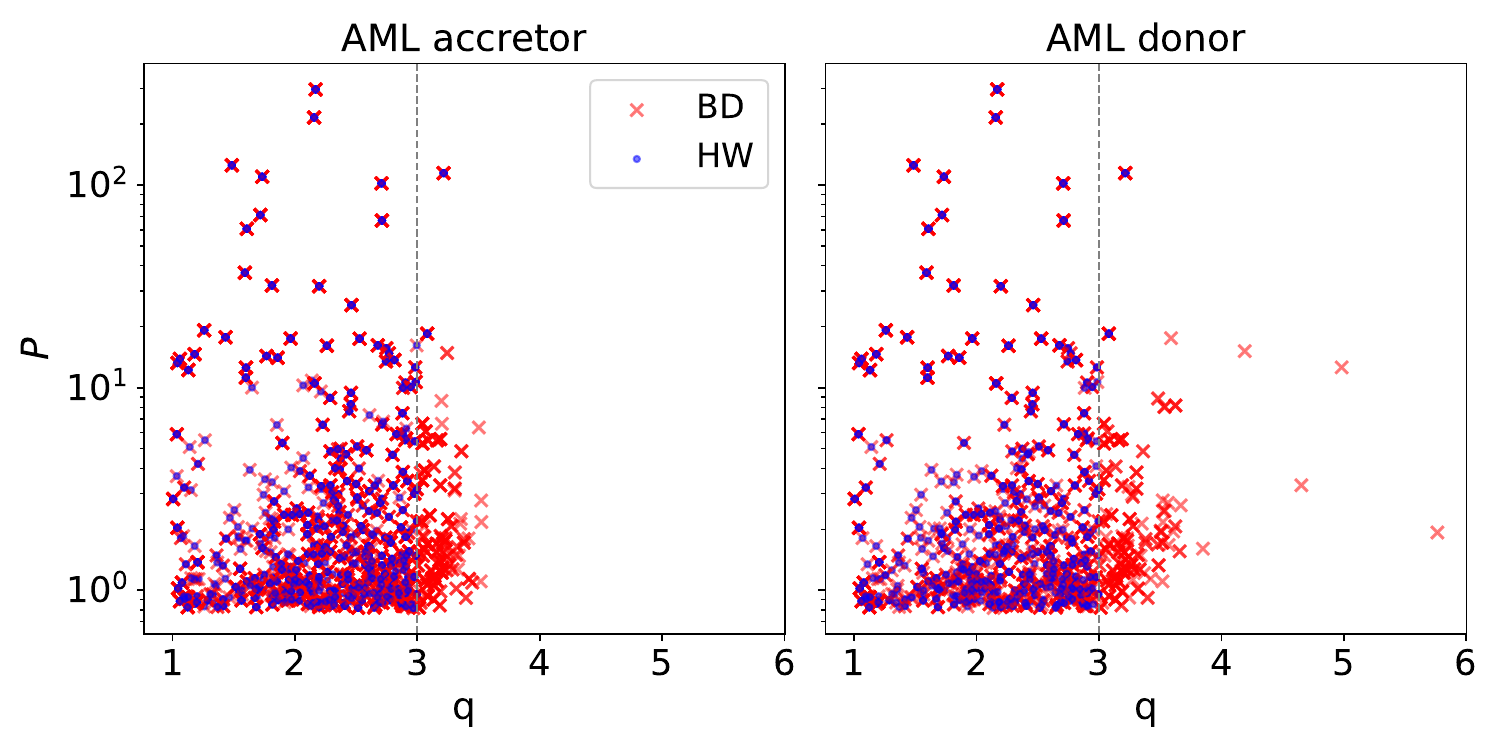}
    \caption{Mass ratio as a function of orbital period prior to the formation of BSSs via mergers at 0.5 Gyr. The left (right) panel corresponds to AML from the accretor (donor) star. The vertical dashed line at $q=3$ corresponds to the upper limit at which MT is considered stable in the HW criterion.}
    \label{fig:Qmergers}
\end{figure}

In the first scenario, all BSSs are formed through a direct merger, i.e., without a prior CE. A small percentage of these mergers originates from collisions, that is, when the sum of the stars’ radii exceeds the periastron distance. In the rest of the cases, the direct merger occurs when the primary MS star fills its Roche lobe and initiates stable MT, causing the secondary star to subsequently fill its own Roche lobe. The system ultimately evolves into a contact binary composed of two MS stars, which in \texttt{SEVN} is assumed to culminate in a stellar merger. Therefore, this formation channel requires a stable MT to occur. Figs. \ref{fig:hist_merge_a} and \ref{fig:hist_merge_d} show the formation efficiency of BSSs produced through this channel. Note that the lighter-colored BD bars are plotted behind the HW bars; thus, when the BD bar is not visible, it indicates that both prescriptions produce the same number of BSSs, while when it is visible, it shows the higher number of BSSs obtained in the BD case. It is shown that there is almost no difference between the number of BSSs produced under the BD prescription and the HW one. This is because the mass ratios at which this formation channel operates are well below 3. This limit comes from the restriction in the maximum mass of the stars imposed by the time they spend in the MS, combined with the minimum mass in the simulations. The only age at which there are differences between the BD and HW criteria is 0.5 Gyr, where a larger number of BSSs are produced under the BD prescription. This results from the fact that at this age donors can be more massive, and therefore BSS progenitors have larger mass ratios. As an illustration, Figure \ref{fig:Qmergers} shows the mass ratio as a function of the orbital period before merger for the BSSs formed under the BD and HW criteria at 0.5 Gyr. The larger number of BSSs obtained with the BD prescription occurs at $q > 3$, which in this scheme corresponds to stable MT. Under the HW criterion, these systems undergo unstable MT, which leads to the disruption of the donor star and no accretion. This outcome follows from the assumption that, because MS stars lack a clear core-envelope separation, an unstable MT episode would cause the entire donor to be rapidly engulfed within the Roche lobe of the other star. Such a rapid event would prevent accretion, leading instead to mass loss through the outer Lagrangian points $L_2$ and $L_3$. This qualitative scenario requires further investigation. It is worth noting that, had we instead assumed that these systems always merge under unstable MT, no differences would arise between the HW and BD criteria.

\begin{figure}
    \centering
    \includegraphics[width=\linewidth]{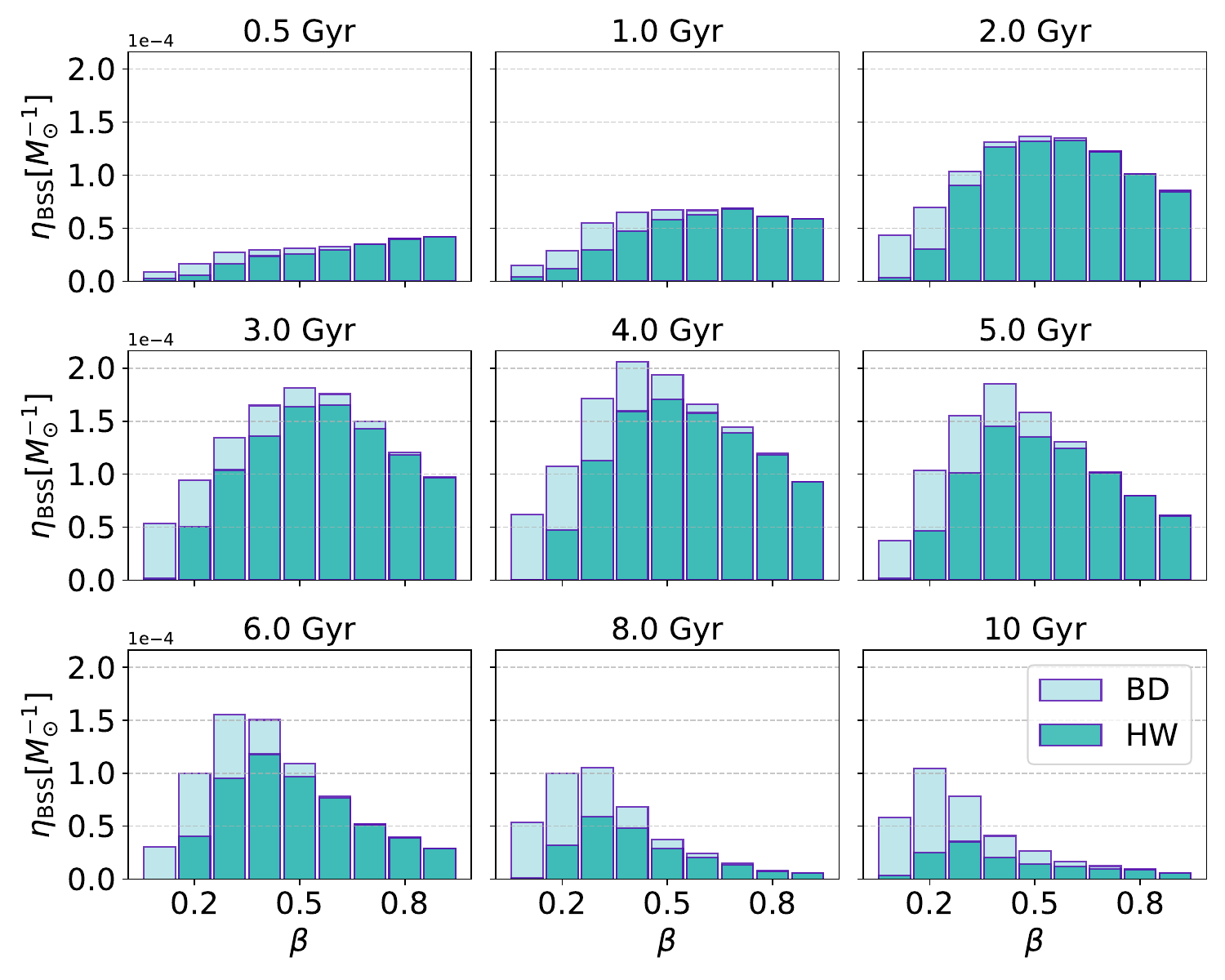}
    \caption{Formation efficiency of BSSs in binary systems (second scenario) as a function of the MT efficiency $\beta$, with AML from the vicinity of the accretor at different ages. The light teal bars indicate results obtained with the BD criterion, whereas the dark teal bars show those from the HW criterion.}
    \label{fig:hist_a}
\end{figure}

\begin{figure}
    \centering
    \includegraphics[width=\linewidth]{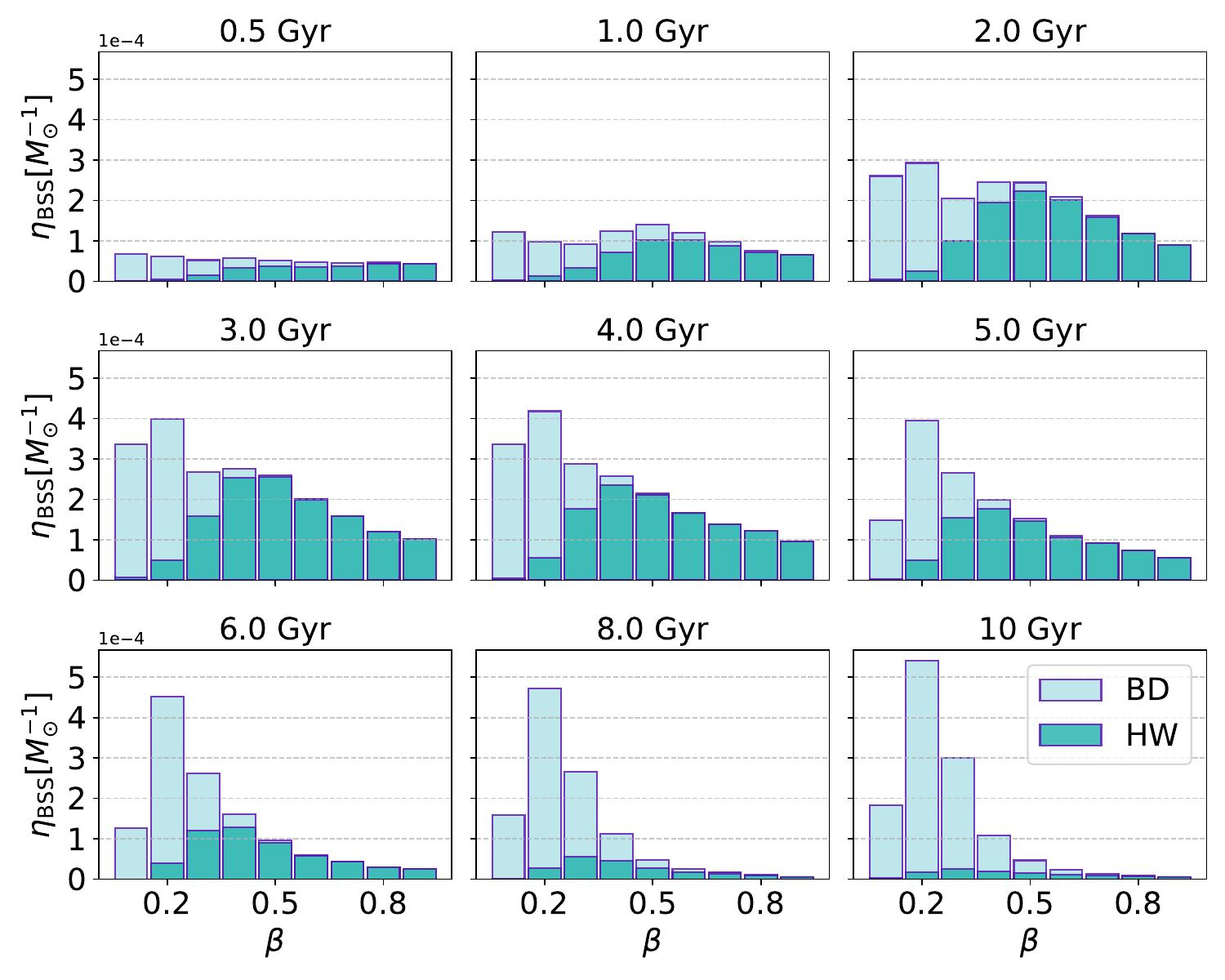}
    \caption{Same as Figure \ref{fig:hist_a} but in the case that AML is from the vicinity of the donor.}
    \label{fig:hist_d}
\end{figure}

Figures \ref{fig:hist_a} and \ref{fig:hist_d} show the impact of the BD prescription on the number of BSSs created in the second scenario at different ages. Figure~\ref{fig:hist_a} corresponds to the case in which angular momentum is lost from the accretor, while Figure~\ref{fig:hist_d} shows the same results when angular momentum is lost from the donor. As seen in both figures, considering the MT efficiency on the stability criterion has the consequence of increasing the amount of BSSs or, in the least favorable cases, leaving it unchanged. When AML occurs from the accretor, the BD criterion produces more BSSs than the HW criteria at low values of $\beta$. The difference gradually decreases at intermediate efficiencies and becomes negligible at high $\beta$. This is simply because the larger the MT efficiency, the more similar the BD and HW prescriptions. Moreover, as shown in Figure~\ref{fig:ZetaGB_a}, the increase in the critical mass ratio $q_{\rm c}$ with decreasing $\beta$ is monotonic; consequently, the enhancement of BSSs increases monotonically as $\beta$ decreases. In contrast, when AML occurs from the donor, the increase in the number of BSSs is much more pronounced at small values of $\beta$. This can be understood because $q_c$ shows a rapidly accelerating growth as $\beta$ decreases, and for a low enough $\beta$, $q_c$ does not even exist, i.e., MT is always stable (see Figure~\ref{fig:ZetaGB_d}). As $\zeta_{\rm ad}$ becomes smaller (representative of giant stars), $q_c$ has a small variation as $\beta$ decreases, i.e., it is similar to $q_c$ in the HW prescription, turning abruptly larger at low values of $\beta$. This variation between the AML modes is reflected in the distribution of the increase in the number of BSSs at different MT efficiencies. The differences arising from the BD and HW prescriptions in this formation scenario are analyzed in more detail below.

\begin{figure*}
    \centering
    \includegraphics[width=1.0\linewidth]{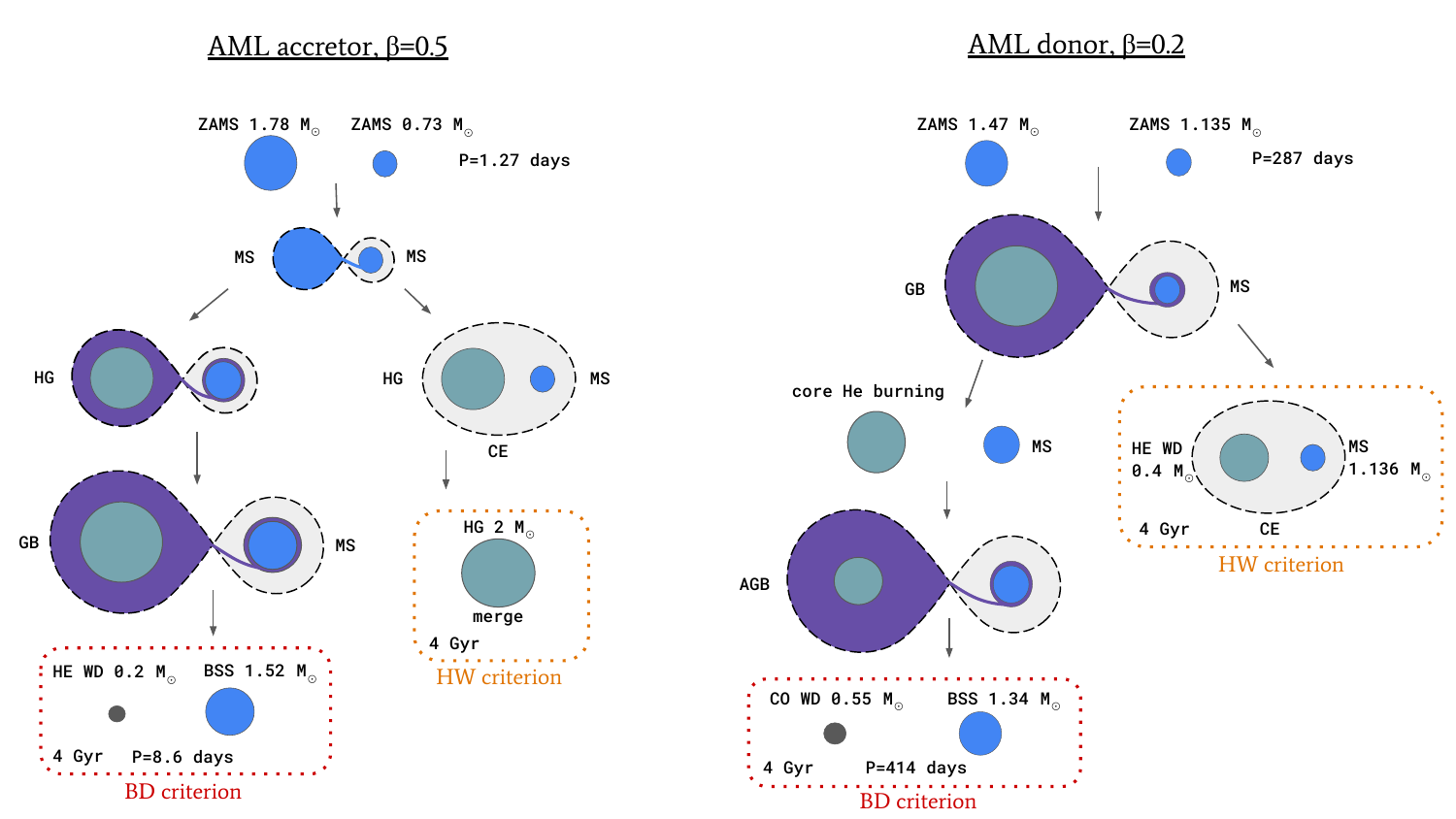}
    \caption{Comparison of the evolutionary history of two binary systems under the BD and HW MT stability criteria. The left side shows the case with AML from the accretor star and MT efficiency $\beta=0.5$, while the right side corresponds to AML from the donor star and $\beta=0.2$. The evolutionary state of the systems at 4 Gyr are highlighted with boxes.}
    \label{fig:esquema}
\end{figure*}

In the second scenario, the primary star fills its Roche lobe and initiates MT onto the secondary MS star. As the donor evolves through more advanced stages, the MT stability is evaluated at every time-step, and the difference between the HW and BD prescriptions emerges, shaping the subsequent evolution of the system. To illustrate this, we consider two representative examples of BSSs formed at 4 Gyr, as shown in Figure \ref{fig:esquema}. The first corresponds to a system initially composed of a $1.78~M_{\odot}$ primary and a $0.73~M_{\odot}$ secondary in a $1.27$~day orbit, evolved with $\beta =$ 0.5 and AML from the accretor. In this system, MT starts with a stable Roche-lobe overflow event when the primary star is in the MS, at the age of $\simeq 1.3$ Gyr. At the age of $\simeq 2.1$ Gyr, the donor star leaves the MS and subsequently evolves to the subgiant and giant branch. With the BD criterion, MT remains stable and ceases at $\simeq 3.9$ Gyr, leaving behind a He WD orbiting a BSS in a $8.6$~day orbit. In contrast, under the HW criterion, the onset of the subgiant phase leads to a CE episode and ultimately a merger. The different fates of these binaries relate to the difference between the BD and HW criteria after the donor star leaves the MS. A second example involves systems with AML from the donor and low MT efficiency. One such case consists of an initial $1.47~M_{\odot}$ primary and a $1.13~M_{\odot}$ secondary in a 287 day orbit, evolved with $\beta = 0.2$. Here, stable MT begins when the donor is on the GB, at $\simeq 2.9$ Gyr. With the BD criterion, the system maintains stable MT until the donor starts burning helium in the core. As the star ascends the asymptotic giant branch and expands further, a second stable MT episode occurs, ending when the donor becomes a carbon-oxygen WD and leaving behind a BSS–WD binary with an orbital period of $\sim 414$ days. However, under the HW criterion, the GB donor cannot sustain stable MT and a CE is formed, in which the accretor cannot gain sufficient mass to exceed the turn-off value of the cluster. These examples demonstrate that accounting for the MT efficiency in the stability criterion can prevent premature CE phases or mergers, allowing systems to undergo stable evolution and produce BSSs in binaries that would otherwise fail to form under the standard HW prescription. 

\begin{figure}
    \centering
    \includegraphics[width=1.0\linewidth]{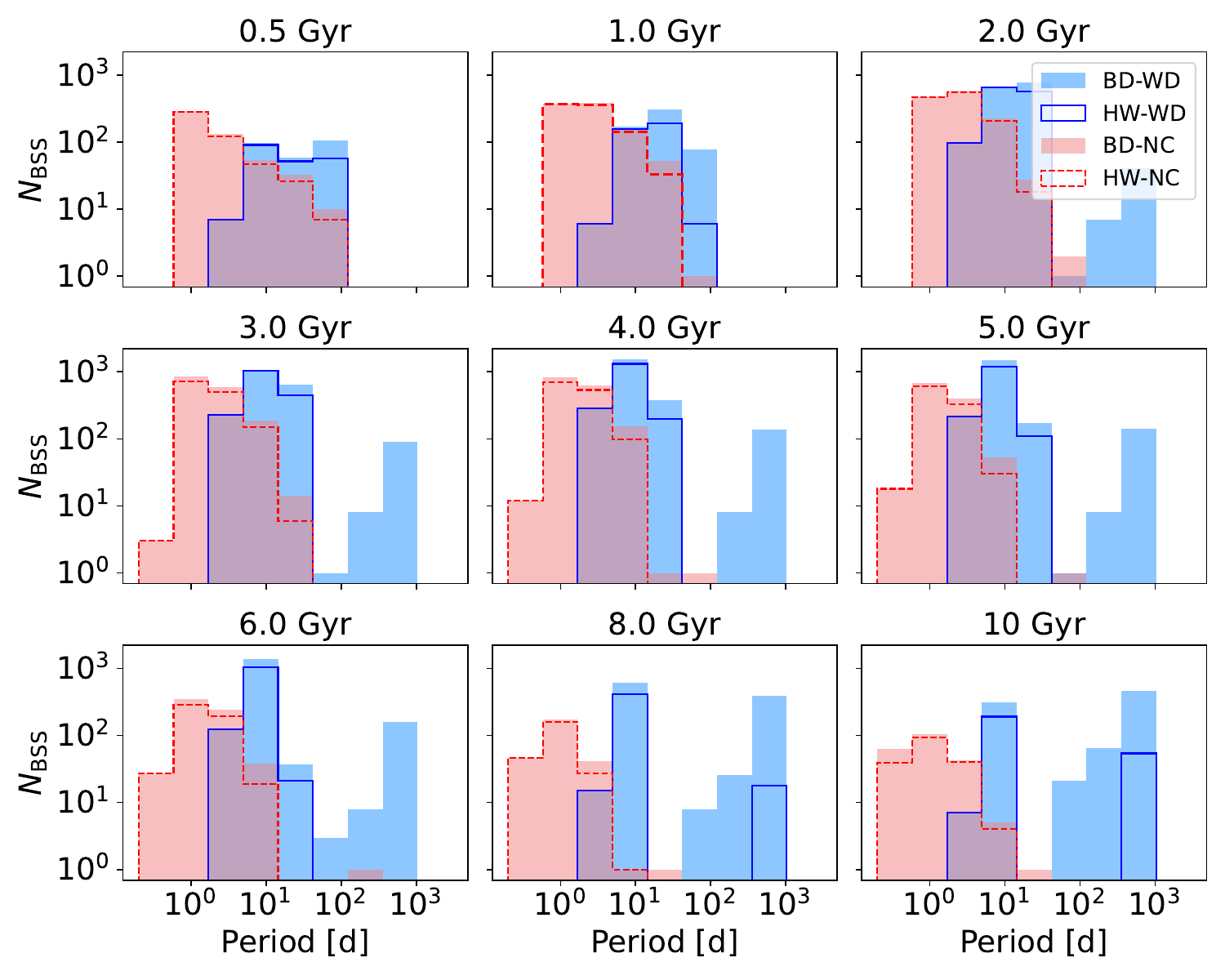}
    \caption{Histogram of the orbital period of BSSs with WD and non-compact (NC) companions, with AML from the vicinity of the accretor at different ages. Models with the BD criterion appear in solid colors: blue for BSSs with WD companions and red for BSSs with NC companions. Models with the HW criterion appear as step-like lines: continuous blue line for BSSs with WD companions and dashed red line for BSSs with NC companions.}
    \label{fig:P_a}
\end{figure}

\begin{figure}
    \centering
    \includegraphics[width=1.0\linewidth]{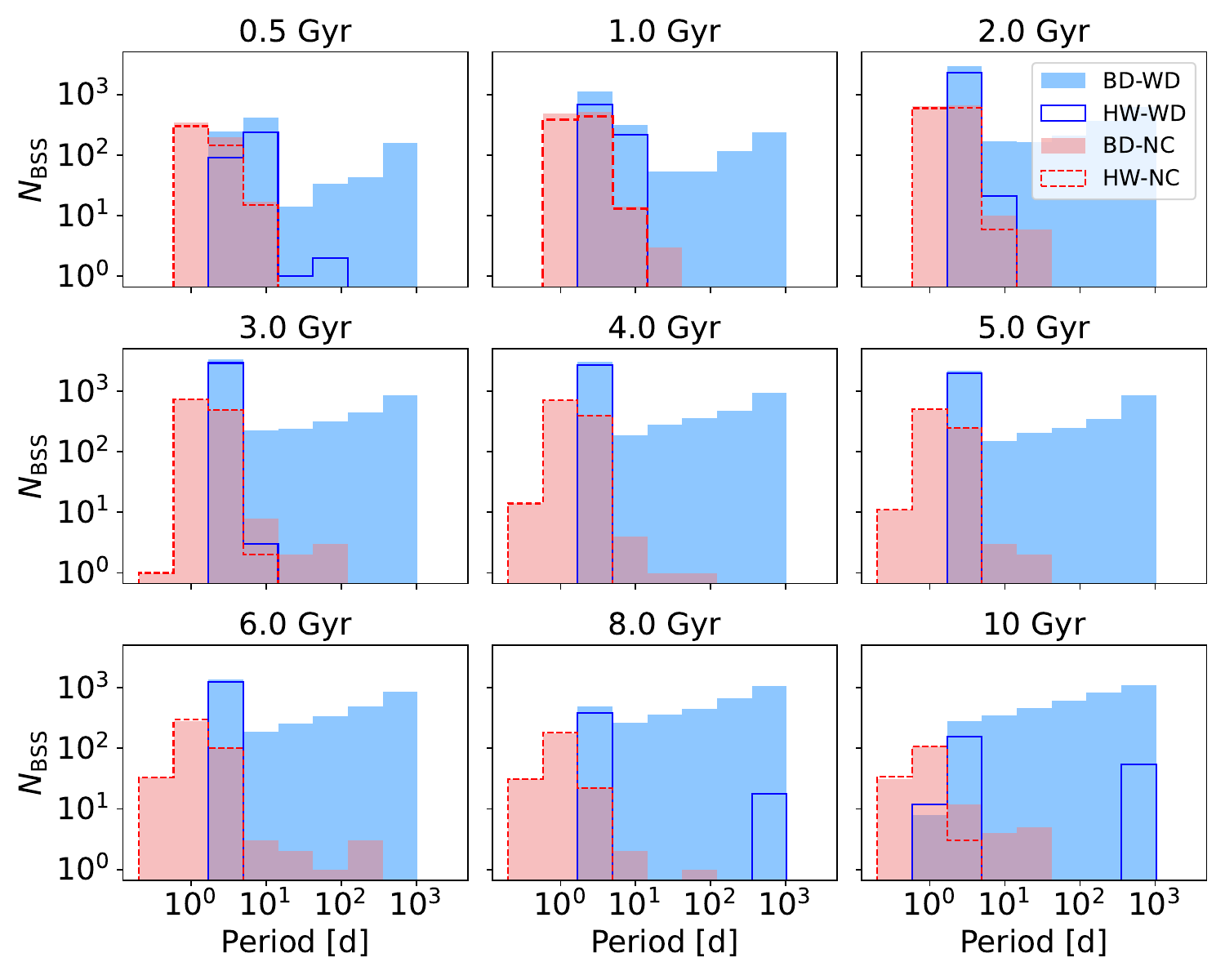}
    \caption{Same as Figure \ref{fig:P_a} but in the case that AML is from the vicinity of the donor.}
    \label{fig:P_d}
\end{figure}

We show the orbital period distribution of the BSSs in Figures \ref{fig:P_a} and \ref{fig:P_d}, distinguishing between those with WD companions (in light blue for the BD criterion and a blue solid line for the HW) from those with non-compact (MS, HG and GB stars) companions (in light red for the BD criterion and a red dashed line for the HW). At each age, we plot the BSSs formed with all values of $\beta$. Compared with the HW prescription, the BD one favors the formation of BSSs in larger orbits, mainly in the case of WD companions. If the AML is from the donor, this tendency is more pronounced, with the longer period binaries being those that undergo highly inefficient MT.

\subsubsection{BSS discussion}

MT in interacting binaries is believed to dominate the formation of BSSs in sparse environments such as open clusters \citep{Rain_2024} and low-density globular clusters \citep{Sollima_2008}, and to be significantly important in denser environments such as cluster cores \citep{Knigge_2009}. In particular, high binary discovery fraction have been found among BSSs in several open clusters \citep{Mathieu_2009,Geller_2011}, with orbital periods ranging from 700 to 3000 days. The companion mass distribution and further far-ultraviolet analysis \citep{Geller_2012,geller2015,Gosnell_2015,pandey2021,Vaidya_2022b} suggest the presence of WD companions to the BSSs, supporting binary interaction where mergers are avoided as the formation channel for these objects. 

Using population synthesis with $N$-body simulations that account for dynamical interactions, \citet{2005MNRAS.363..293H} and \citet{Geller_2012} studied BSSs in the open clusters M67 ($\sim 4$ Gyr old) and NGC 188 ($\sim 7$ Gyr old) and found a discrepancy between the number of objects formed in the simulations and the number of BSSs observed, the simulated ones being about a third of those observed. \citet{Tian_2006} also studied the formation of BSSs in M67 and predicted only a few objects formed via stable MT, far fewer than observed. In addition, their models have only very short orbital periods (less than 10 days). \citet{Leiner_2021} studied 16 open clusters with ages ranging from 1 to 10 Gyr and showed that simulations performed with the \texttt{COSMIC} rapid binary-evolution code dramatically under-produce the number of BSSs. They concluded that neither mergers nor collisions can fully account for these differences and suggested that MT from giant donors must be more stable than typically assumed. Furthermore, short-period binaries are easily reproduced by population synthesis, but longer period systems are remarkably difficult to form \citep{nie2012MNRAS.423.2764N, Leiner_2021}. Here, we have shown that a more consistent treatment of MT stability has the effect of increasing the number of BSSs at many ages and MT efficiencies, while also favoring their formation in wider orbits. These results suggest that this approach may help mitigate some of the discrepancies between theoretical predictions and observations.

\subsection{Binary compact objects}
\label{sec:BCO}

Here we focus on BCOs composed of two black holes (BHs), two neutron stars (NSs), or a BH and a NS. They are potential progenitors of observed GW events if they merge within the age of the Universe (e.g., \citealt{abbottPhysRevX.9.031040, Abbott_2020, abbottPhysRevX.11.021053, Zackay2019PhysRevD.100.023007, Nitz_2020, Venumadhav2020PhysRevD.101.083030}). The time-lapse between the formation of the binary at the ZAMS and the merger event is referred to as the delay time. A system is defined as a GW progenitor when its delay time is less than the age of the Universe, for which we employ a value of $13.8$~Gyr as a reference \citep[e.g.,][]{Planck2020}.

BCOs can form through isolated binary evolution, in which massive binaries remain bound until both compact objects are formed. The different evolutionary pathways that lead to the formation of BCOs in this scenario are typically classified according to the type of binary interactions that occur before and after the formation of the first compact object. Following the classification by \citet{Iorio2023} and \citet{Broekgaarden2021MNRAS.508.5028B}, we define the following channels. Channel I: The system interacts only through stable MT before the formation of the first compact object, after which there is at least one CE phase. Channel II: The system evolves only through stable MT episodes. Channel III: Before the formation of the first compact object, the system goes through at least one CE. In addition, at the moment of the first compact object formation, the system is composed of one H-rich star and one star without H envelope. Channel IV: Similar to channel III but, at the time of the first compact object formation, both stars have lost their H envelope. Channel V: There are no interactions before the formation of the first compact object. Channel 0: There are no interactions at all. 

\begin{figure*}h
    \centering
    \includegraphics[width=\linewidth]{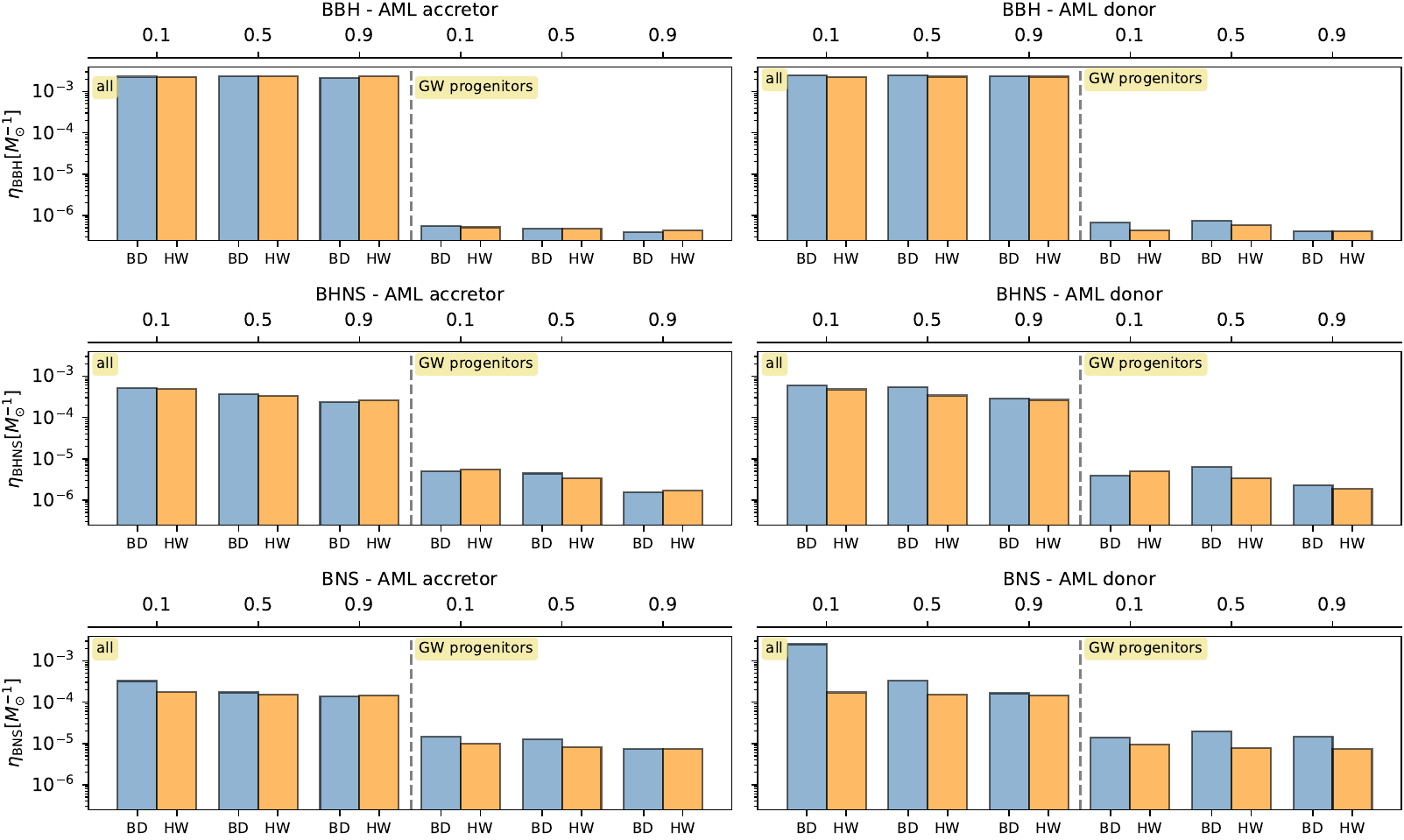}
    \caption{Formation efficiency of BBH, BHNS, and BNS systems for different MT efficiencies ($\beta = 0.1, 0.5, 0.9$) and stability prescriptions (BD and HW). The left and right columns correspond to AML from the vicinity of the accretor and the donor, respectively. In each panel, all BCO systems are shown to the left of the vertical dashed line, while GW progenitors are shown to the right.}
    \label{fig:histBCO}
\end{figure*}

In this section, we show the effects of the new BD prescription on the formation of BCO and GW progenitors, distinguishing between binary BHs (BBHs), binary NSs (BNSs), and BH-NS binaries (BHNSs). The sample we produced for the analysis consists of $10^{7}$ binaries per each $\beta$ = $0.1$, $0.5$, and $0.9$, each MT prescription (HW and BD), and each AML mode (from the vicinity of the donor or the accretor star), for a total of 12 sets of simulations. Each set is constructed following the \citet{Kroupa2001} initial mass function from $5$ to $150~M_{\odot}$, and the \citet{Sana2012} prescriptions for the period, eccentricity and mass-ratio at the ZAMS, which are based on observations of high-mass O-type stars. Stars are initialized without rotation and in all cases we have adopted a metallicity Z=0.014. At this metallicity, \citet{Iorio2023} have shown that channel I is inefficient in producing BBH GW progenitors, while channel II provides a non-negligible contribution. On the other hand, for BHNS GW progenitors, channel I has a small contribution and channel II is negligible, while both channels are suppressed for BNS GW progenitors. This implies that, in the HW prescription, channels involving unstable mass transfer (i.e., channels I, III, and IV) play an important role in the formation of BCO GW progenitors. Therefore, this metallicity regime provides a good test case to evaluate the impact of the new formalism, which tends to favor more stable MT episodes.

The initial conditions account for a total simulated mass of $2.18 \times 10^{7}~M_{\odot}$, which corresponds to an effective total mass $M_{\rm pop}=8.55 \times 10^{7}~M_{\odot}$.\footnote{We computed this quantity taking into account the incompleteness of the simulated IMF, period, and mass-ratio distributions, as well as assuming a binary fraction of 1. The fraction of simulated mass in this case is $0.255$.} The same list of initial conditions has been used in all cases, including the adoption of the same seed number for the supernova kick velocities (since it is randomly sampled from a Maxwellian distribution), in order to measure the differences caused only by the MT stability prescription, the AML mode, and the value of $\beta$. The evolution of the binaries is simulated until the formation of the BCO, if it happens\footnote{For the remaining cases, the stopping conditions are the disruption of the binary and the evolution of a merger product until the formation of a compact object.}. When a BCO is formed, the time from its formation to the GW emission (which we refer to as the fusion time) is calculated following the high-precision analytic approximation which can be found in Appendix D in \citet{Iorio2023}. Thus, the delay time is computed as the time span between the ZAMS and the formation of the BCO plus the fusion time.

\subsubsection{BCO results}

Figure~\ref{fig:histBCO} shows the formation efficiency $\eta_{\rm BCO}=N_{\rm BCO}/M_{\rm pop}$, with $N_{\rm BCO}$ the number of each BCO type in the simulation. The top, middle, and bottom row corresponds to BBH, BHNS, and BNS systems, respectively. We consider angular momentum loss from the accretor and the donor, three values of MT efficiency, and two stability prescriptions (BD and HW). The vertical gray line separates the full BCO population from the subset of systems classified as GW progenitors. For BBHs and BHNSs, the formation efficiencies obtained with the BD and HW prescriptions are largely similar, both for the general populations and for the GW progenitors. In contrast, BNSs exhibit a generally higher formation efficiency under the BD prescription at low values of $\beta$, with the largest differences occurring when angular momentum is lost from the donor. A similar trend is observed for BNS GW progenitors, which are systematically produced more efficiently in the BD case than in the HW one.

\begin{figure*} 
    \centering
    \includegraphics[width=\linewidth]{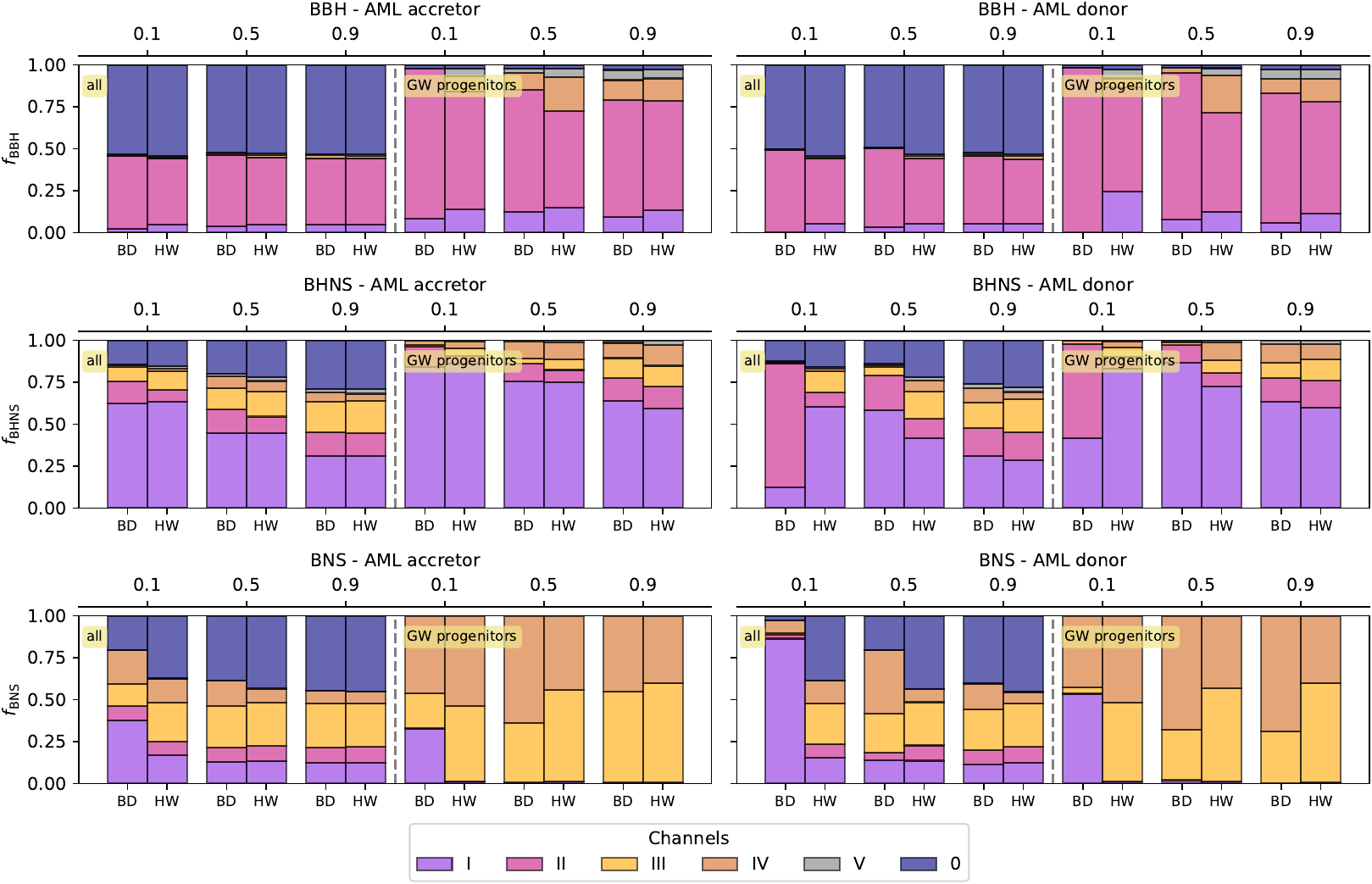}
    \caption{Relative fractions of BBH, BHNS, and BNS systems across formation channels (see Sect.~\ref{sec:BCO}), for different MT efficiencies ($\beta = 0.1, 0.5, 0.9$) and stability prescriptions (BD and HW). The left and right columns correspond to angular momentum loss from the vicinity of the accretor and the donor, respectively. In each panel, all BCO systems are shown to the left of the vertical dashed line, while GW progenitors are shown to the right.}
    \label{fig:canales}
\end{figure*}

Figure~\ref{fig:canales} shows the relative fractions of the formation channels described in Sect.~\ref{sec:BCO} for BBH, BHNS, and BNS systems and the corresponding subset of GW progenitors. For each BCO type, the relative fraction of a given channel is defined as the number of systems formed through that channel divided by the total number of systems. For GW progenitors, the relative fraction is instead normalized to the total number of GW progenitors. In the following, we refer to these relative fractions as channel efficiencies.

For BBHs, the channel efficiencies obtained with the BD and HW prescriptions are largely similar, with channels II and 0 dominating in all cases. The main difference is a modest increase in the efficiency of channel II under the BD prescription, particularly when angular momentum is lost from the donor. For BBH GW progenitors, channel II remains the dominant formation pathway, followed by channel I and, at higher MT efficiencies, channel IV. In this case, the increase in the efficiency of channel II under the BD prescription becomes more pronounced.

In the case of BHNS, when angular momentum is lost from the accretor, channel I is the dominant formation pathway under both the HW and BD prescriptions, and the adoption of the BD criterion mainly results in a moderate increased efficiency of channel II. When AML proceeds from the donor, channel I dominates under the HW prescription for all $\beta$ values. For $\beta = 0.5$, both channels I and II become more efficient with the BD prescription, and channel I is still the dominant formation channel. Instead, for highly non-conservative MT ($\beta = 0.1$), channel II becomes dominant, increasing from 0.08 to 0.74, while channel I is strongly suppressed and channel III disappears. A similar behavior is found for BHNS GW progenitors. When AML is from the accretor, channels involving stable MT (I and II) generally increase at the expense of CE-dominated channels when considering the BD prescription. If AML is from the donor, channel I remains dominant at $\beta = 0.5$ and 0.9, while channel II becomes more efficient at $\beta = 0.1$ than channel I.

For BNS, channels III and IV are more important than for the BHBH and BHNS systems, reflecting the central role of CE evolution in the production of these systems. For intermediate and high MT efficiencies, changing the stability prescription does not lead to substantial differences, except for the case with $\beta = 0.5$ and AML from the donor, where the efficiency of channel IV increases at the expense of channel 0. In contrast, for highly non-conservative MT ($\beta = 0.1$), the increased stability of MT significantly alters the channel efficiency. In this regime, the efficiency of channel I rises markedly, more than doubling when AML is from the accretor and increasing from 0.15 to 0.86 when AML is from the donor. A similar trend is found for BNS GW progenitors, with the difference that channel 0 is completely suppressed. While channels III and IV are the main contributors for intermediate and high $\beta$, channel I becomes more important at $\beta = 0.1$, with its efficiency rising from 0.01 to 0.32 if AML proceeds from the accretor and to 0.53 if AML is from the donor.

\begin{figure}
    \centering
    \includegraphics[width=\linewidth]{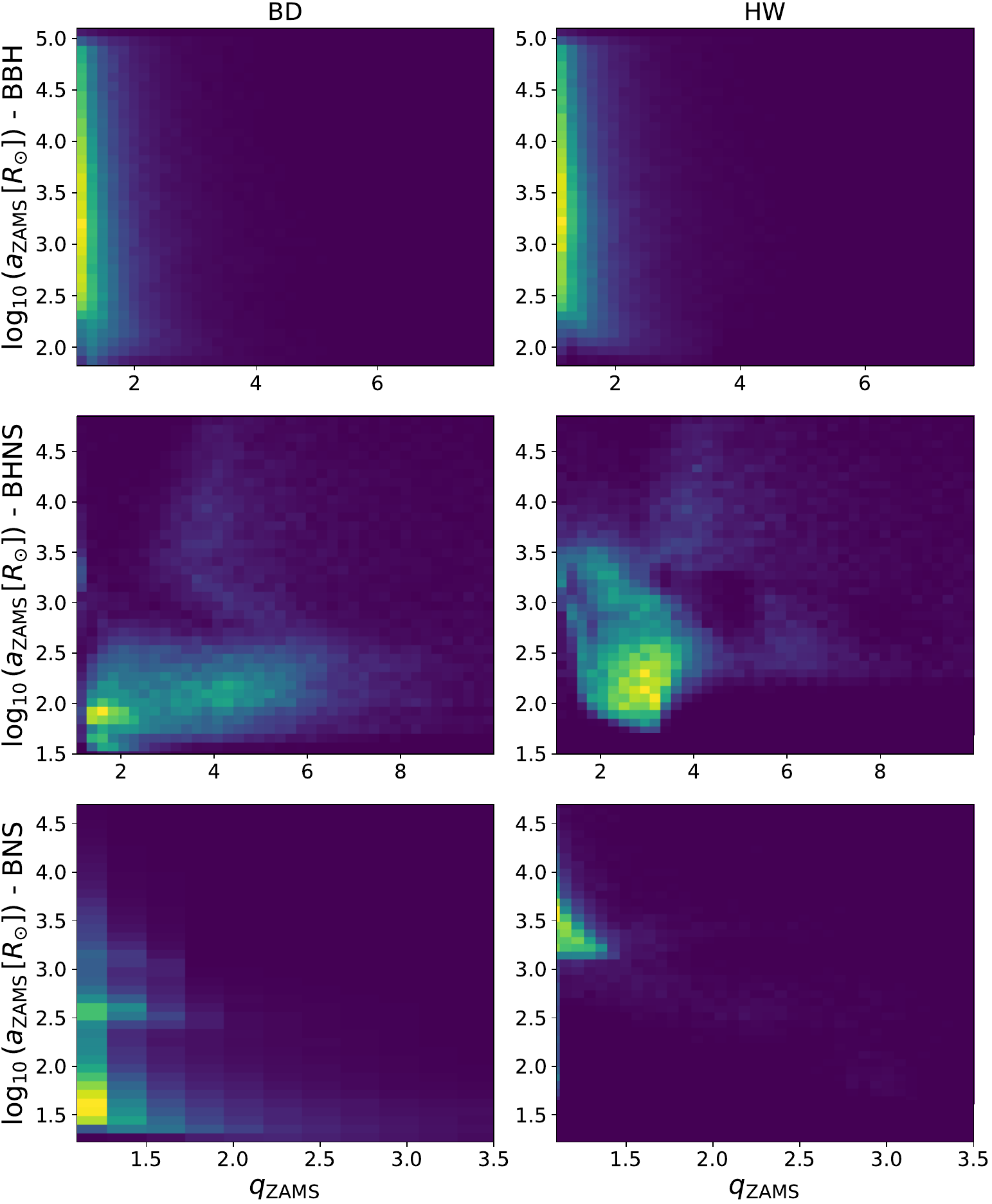}
    \caption{Mass-ratio and semi-major axis joint distributions at ZAMS for BBH (top panel), BHNS (middle panel), and BNS (bottom panel). Left panel: BD prescription. Right panel: HW prescription. The color scale represents the joint probability density distribution, normalized such that the integral over each panel is unity. Colors range from low (dark) to high (light) density.}
    \label{fig:ICs_comparison}
\end{figure}

Figure~\ref{fig:ICs_comparison} shows the ZAMS mass ratio, $q_{\rm ZAMS}$, and semi-major axis, $a_{\rm ZAMS}$, of BBH, BHNS, and BNS progenitors formed under the BD and HW prescriptions, for the case $\beta = 0.1$ and AML from the donor. We show this case because it is the one in which the differences between the BD and HW prescriptions are most pronounced. For BBHs, we have shown that the efficiency of channel II increases, while the total number of systems remains essentially unchanged (showing a moderate increase of GW progenitors in the case of the BD prescription). This is reflected in the distribution of $q_{\rm ZAMS}$ and $a_{\rm ZAMS}$, which differ especially at small values of $a_{\rm ZAMS}$, where the BD prescription produces a larger number of BBH GW progenitors. In the case of BHNS progenitors, the BD prescription leads to the substitution of channel I with channel II as the main formation channel, and an overall increase in the number of systems formed. This is accompanied by a shift of the distribution maximum toward lower values of $q_{\rm ZAMS}$ and $a_{\rm ZAMS}$. In addition, the region of the parameter space above $\log (a_{\rm ZAMS}/ R_{\odot}) \simeq 2.7$ that is populated under the HW prescription becomes nearly devoid of progenitors when the BD criterion is adopted. This indicates that systems which would form BHNS progenitors under the HW prescription instead evolve into different types of binaries when the BD stability criterion is applied. For BNS progenitors, we have shown that the BD prescription produces an even more dramatic reorganization of the formation channels. Channels III and 0 are largely replaced by channel I as the dominant formation channel, leading to a pronounced increase in the total number of systems formed. This transition is accompanied by a substantial change in the distribution of initial conditions. Under the HW prescription, most BNS progenitors originate from systems with $\log (a_{\rm ZAMS} / R_{\odot}) \simeq 3.5$, whereas in the BD case, progenitors are preferentially drawn from orbits roughly two orders of magnitude tighter. The strong concentration toward low values of $a_{\rm ZAMS}$ naturally explains the enhanced formation efficiency of BNSs in the BD scenario, as the initial condition distributions populate this region more densely. As in the BHNS case, but more markedly, many systems that form BNS progenitors under the HW prescription instead evolve into different outcomes when the BD criterion is adopted.

\subsubsection{BCO discussion}

Several studies investigate the relative contribution of different formation channels to the formation of BCOs. Channels involving at least one CE phase have previously been identified as the dominant evolutionary pathways to form BBH (e.g., \citealt{bavera2021A&A...647A.153B, ramon2021ApJ...912L..23R}). However, \citet{Olejak2021} found a significantly larger contribution from stable MT channels, and explicitly showed that the relative importance of the different channels is highly sensitive to the adopted MT stability criteria, even finding the emergence of a dominant formation scenario for BBH GW progenitors without a CE phase. Similarly, \citet{dorozsmai2024MNRAS.530.3706D} identified the stability of MT as a key factor in increasing the relative importance of stable MT channels, while \citet{Broekgaarden2021MNRAS.508.5028B} found that channel II dominates over channel I for BHNS GW progenitors, a result that can be traced back to the adopted stability prescriptions and MT efficiencies tested in their work. 

There is also an ongoing tension between theoretical predictions and GW observations, where MT stability contributes as one of the main sources of uncertainty. Several studies have suggested that population synthesis models may overestimate the occurrence of CE events, or alternatively, that stable MT may play a more prominent role in BBH formation than previously expected (e.g., \citealt{Pavlovskii2017MNRAS.465.2092P,giacobbo2018MNRAS.474.2959G,Neijssel2019MNRAS.490.3740N,Olejak2021,bavera2021A&A...647A.153B,Marchant2021,vanSon2022ApJ...931...17V}), and that this could partially contribute to the apparent excess of merging BBHs predicted by these models when compared with GW merger rates inferred by observations (e.g., \citealt{Boesky2024ApJ...976...24B,Sgalletta2025, Marinacci2025}). Moreover, detailed stellar structure calculations tend to favor stable MT (e.g., \citealt{ge2010ApJ...717..724G,Ge2015ApJ...812...40G,Ge2020ApJ...899..132G,Ge2024ApJ...975..254G}), reinforcing the idea that CE events may be over-predicted on population synthesis calculations. Recently, \citet{li2025PhRvD.112j3005L} used \texttt{MOBSE} simulations that include chemically homogeneous evolution and found that the low-mass peak of the observed BBH mass distribution ($\sim 10~M_{\odot}$) can be reproduced through either CE evolution or stable MT depending on the adopted stability criterion, while the high-mass peak ($\sim 35~M_{\odot}$) is attributed to the chemically homogeneous evolution channel. Notably, their predicted merger rates span nearly a factor of six across models, underscoring how sensitively population-level observables respond to these choices. In this context, our finding that the BD formalism systematically shifts the balance toward stable MT channels reinforces the need for a careful treatment of MT stability in population synthesis studies, and motivates future work with \texttt{SEVN} aimed at characterizing how this stability shift propagates into the BBH merger-rate density as a function of BH mass. Together, these results highlight the central role of the treatment of MT stability.

Our results support this emerging picture by showing an enhanced efficiency of formation channels involving stable MT when adopting a more self-consistent stability criterion. By increasing the parameter space in which MT remains stable, our prescription naturally favors channels dominated by stable MT, bringing population synthesis predictions closer to those obtained from detailed stellar-evolution calculations and potentially helping to alleviate existing discrepancies with observational constraints.

\section{Summary and conclusions}
\label{sec:conclusion}

The primary goal of this work is to investigate the impact of a more self-consistent treatment of MT stability. To this end, we have developed a new formalism for the stability of MT through Roche-lobe overflow that explicitly accounts for the loss of mass and angular momentum. In this first implementation, we consider two modes of angular momentum loss, namely from the vicinity of the accretor or the donor star, such that the resulting stability criterion depends on the mass ratio of the binary and on the efficiency of the mass transfer, which is parametrized by $\beta$. This new $\beta$-dependent stability criterion enables systems with larger mass ratios to undergo stable MT, as the critical mass ratio below which MT is stable increases as $\beta$ decreases. We have implemented this new formalism on the population synthesis code \texttt{SEVN} and used BSSs and BCOs as representative test cases to exemplify the impact of the new prescription on the evolution of interacting binaries.

In the case of BSSs, the new stability criterion enhances their formation efficiency and favors the production of systems in wider orbits compared to the standard stability formalism. This behavior arises because, in our simulations, BSSs form predominantly through stable MT. In particular, systems with the widest orbits are those that undergo inefficient but stable MT while the donor is on the GB. These results point in the right direction, since previous studies have reported discrepancies between population synthesis predictions and observations, including an underproduction of BSSs and difficulties in forming systems with long orbital periods. 

In the case of BCOs, the effect of the new stability prescription varies between different types of systems. This reflects the fact that the relative importance of the formation channels involving CE evolution and stable MT differs among BCO populations. The most significant differences are found in binaries containing at least one NS, especially for low values of $\beta$ and when angular momentum is lost from the donor, where the new criterion leads to higher formation efficiencies compared to the standard prescription. Our results show that the BD prescription generally favors evolutionary channels involving stable MT. In particular, the BD prescription favors channel II for BBHs and BHNSs, and favors channel I for BNSs. The changes are more pronounced at low MT efficiency and AML from the donor, where the formation of GW progenitors is also affected. These results should, however, be interpreted in the context of the broader parameter space governing BCO formation, which includes additional uncertainties such as supernova kicks and CE parameters, and therefore they represent a first, non-exhaustive exploration. In particular, recent studies have provided improved prescriptions for the envelope binding energy \citep{xu2010ApJ...716..114X,Loveridge2011ApJ...743...49L} and tighter constraints on the CE ejection efficiency \citep{Zorotovic2022MNRAS.513.3587Z,Scherbak2023MNRAS.518.3966S, ge2022ApJ...933..137G, ge2024ApJ...961..202G} based on detailed stellar evolution models, both of which may significantly affect the relative importance of different formation channels. In addition, there is a strong dependence of channel efficiencies on metallicity (see, e.g., \citealt{Klencki2020A&A...638A..55K,Klencki2021A&A...645A..54K,Iorio2023}). Our new formalism, which will be soon available in the public code \texttt{SEVN,} will make it possible to carry out comprehensive population studies of BCOs, allowing us to explore the impact of MT efficiency across an extended parameter space. 

In future work, we plan to further refine the description of Roche-lobe overflow MT by incorporating a more detailed treatment of the donor’s response to mass loss. As discussed in Sect.~\ref{sec:2}, the stability of MT critically depends on the adiabatic response of the donor, $\zeta_{\rm ad}$, which in this work has been approximated by fixed values for each evolutionary phase. A more realistic description of this response, computed at solar metallicity and provided in tabulated form by \citet{Ge2015ApJ...812...40G,Ge2020ApJ...899..132G,Ge2023ApJ...945....7G,Ge2024ApJ...975..254G}, would naturally complement the $\beta$-dependent formalism for $\zeta_L$ presented here, leading to a more self-consistent criterion for MT stability. 

In addition, the efficiency of MT and the way in which angular momentum is lost from the system remain major uncertainties, both of which have a strong impact on the binary evolution. A crucial aspect concerns the amount of material and angular momentum the accretor can gain when it approaches its critical rotation. Some studies suggest that if tidal forces are strong enough, the accretor can spin down, enabling a higher MT efficiency. Within this framework, short-period binaries would experience higher MT efficiency, while long-period systems would undergo highly non-conservative MT \citep{Wang2022NatAs...6..480W}. Furthermore, the interaction between the accretor and the accretion disk may allow accretion to continue beyond the critical rotation limit by transporting angular momentum outward through the disk \citep{Popham1991}. Ultimately, relaxing the assumption of fixed MT efficiency and exploring its dependence on the orbital period would represent a natural next step toward a more realistic description of Roche-lobe overflow MT.

In this work, we have considered two modes of AML, namely isotropic re-emission and the Jeans mode. However, \texttt{SEVN} also includes a third prescription in which the ejected material forms a circumbinary disc and extracts additional angular momentum from the system, potentially leading to further modifications of the stability boundaries derived here. The impact of this mechanism on the stability of MT and on the resulting binary populations remains to be explored.

In conclusion, we have shown that a more self-consistent treatment of the MT stability criterion can have a significant impact on binary population synthesis results, affecting the formation and properties of both  BSSs and BCOs.

\begin{acknowledgements}

{\bf Funding.} 
Funded by the European Union (ERC-2022-AdG, {\em "StarDance: the non-canonical evolution of stars in clusters"}, Grant Agreement 101093572, PI: E. Pancino). Views and opinions expressed are however those of the author(s) only and do not necessarily reflect those of the European Union or the European Research Council. Neither the European Union nor the granting authority can be held responsible for them. GJE akncowledges support from the Spanish Ministry of Science via the Plan de Generación de Conocimiento through grant PID2022-143331NB-100. GI is supported by a fellowship grant from la Caixa Foundation (ID 100010434). The fellowship code is LCF/BQ/PI24/12040020. For the \texttt{SEVN} development, GJE, GI, and MP acknowledge financial support from the European Research Council for the ERC Consolidator grant DEMOBLACK, under contract no. 770017, PI: M. Mapelli. EL acknowledges support from the ERC Consolidator Grant funding scheme (project ASTEROCHRONOMETRY, G.A. n. 772293). We thank the referee, Hongwei Ge, for the careful reading of our manuscript and for the constructive comments and suggestions.
\\

\end{acknowledgements}

\bibliographystyle{aa} 
\bibliography{biblio}

\begin{thebibliography}{136}
\expandafter\ifx\csname natexlab\endcsname\relax\def\natexlab#1{#1}\fi

\bibitem[{Abbott {et~al.}(2020)Abbott, Abbott, Abbott, Abraham, Acernese,
  Ackley, Adams, Adhikari, Adya, Affeldt, Agathos, Agatsuma, Aggarwal, Aguiar,
  Aiello, Ain, Ajith, Allen, Allocca, Aloy, Altin, Amato, Anand, Ananyeva,
  Anderson, Anderson, Angelova, Antier, Appert, Arai, Araya, Areeda, Arène,
  Arnaud, Aronson, Arun, Ascenzi, Ashton, Aston, Astone, Aubin, Aufmuth,
  AultONeal, Austin, Avendano, Avila-Alvarez, Babak, Bacon, Badaracco, Bader,
  Bae, Baird, Baker, Baldaccini, Ballardin, Ballmer, Bals, Banagiri, Barayoga,
  Barbieri, Barclay, Barish, Barker, Barkett, Barnum, Barone, Barr, Barsotti,
  Barsuglia, Barta, Bartlett, Bartos, Bassiri, Basti, Bawaj, Bayley, Baylor,
  Bazzan, Bécsy, Bejger, Belahcene, Bell, Beniwal, Benjamin, Berger, Bergmann,
  Bernuzzi, Berry, Bersanetti, Bertolini, Betzwieser, Bhandare, Bidler, Biggs,
  Bilenko, Bilgili, Billingsley, Birney, Birnholtz, Biscans, Bischi,
  Biscoveanu, Bisht, Bitossi, Bizouard, Blackburn, Blackman, Blair, Blair,
  Blair, Bloemen, Bobba, Bode, Boer, Boetzel, Bogaert, Bondu, Bonnand, Booker,
  Boom, Bork, Boschi, Bose, Bossilkov, Bosveld, Bouffanais, Bozzi, Bradaschia,
  Brady, Bramley, Branchesi, Brau, Breschi, Briant, Briggs, Brighenti, Brillet,
  Brinkmann, Brockill, Brooks, Brooks, Brown, Brunett, Buikema, Bulik, Bulten,
  Buonanno, Buskulic, Buy, Byer, Cabero, Cadonati, Cagnoli, Cahillane,
  Bustillo, Callister, Calloni, Camp, Campbell, Canepa, Cannon, Cao, Cao,
  Carapella, Carbognani, Caride, Carney, Carullo, Diaz, Casentini, Caudill,
  Cavaglià, Cavalier, Cavalieri, Cella, Cerdá-Durán, Cesarini, Chaibi,
  Chakravarti, Chamberlin, Chan, Chao, Charlton, Chase, Chassande-Mottin,
  Chatterjee, Chaturvedi, Chatziioannou, Cheeseboro, Chen, Chen, Chen, Cheng,
  Cheong, Chia, Chiadini, Chincarini, Chiummo, Cho, Cho, Cho, Christensen, Chu,
  Chua, Chung, Chung, Ciani, Cieślar, Ciobanu, Ciolfi, Cipriano, Cirone,
  Clara, Clark, Clearwater, Cleva, Coccia, Cohadon, Cohen, Colleoni, Collette,
  Collins, Colpi, Cominsky, Constancio, Conti, Cooper, Corban, Corbitt,
  Cordero-Carrión, Corezzi, Corley, Cornish, Corre, Corsi, Cortese, Costa,
  Cotesta, Coughlin, Coughlin, Coulon, Countryman, Couvares, Covas, Cowan,
  Coward, Cowart, Coyne, Coyne, Creighton, Creighton, Cripe, Croquette,
  Crowder, Cullen, Cumming, Cunningham, Cuoco, Canton, Dálya, D’Angelo,
  Danilishin, D’Antonio, Danzmann, Dasgupta, Costa, Datrier, Dattilo, Dave,
  Davier, Davis, Daw, DeBra, Deenadayalan, Degallaix, Laurentis, Deléglise,
  Lillo, Pozzo, DeMarchi, Demos, Dent, Pietri, Rosa, Rossi, DeSalvo, Varona,
  Dhurandhar, Díaz, Dietrich, Fiore, DiFronzo, Giorgio, Giovanni, Giovanni,
  Girolamo, Lieto, Ding, Pace, Palma, Renzo, Divakarla, Dmitriev, Doctor,
  Donovan, Dooley, Doravari, Dorrington, Downes, Drago, Driggers, Du, Ducoin,
  Dudi, Dupej, Durante, Dwyer, Easter, Eddolls, Edo, Effler, Ehrens, Eichholz,
  Eikenberry, Eisenmann, Eisenstein, Errico, Essick, Estelles, Estevez,
  Etienne, Etzel, Evans, Evans, Fafone, Fairhurst, Fan, Farinon, Farr, Farr,
  Fauchon-Jones, Favata, Fays, Fazio, Fee, Feicht, Fejer, Feng,
  Fernandez-Galiana, Ferrante, Ferreira, Ferreira, Fidecaro, Fiori, Fiorucci,
  Fishbach, Fisher, Fishner, Fittipaldi, Fitz-Axen, Fiumara, Flaminio,
  Fletcher, Floden, Flynn, Fong, Font, Forsyth, Fournier, Vivanco, Frasca,
  Frasconi, Frei, Freise, Frey, Frey, Fritschel, Frolov, Fronzè, Fulda, Fyffe,
  Gabbard, Gadre, Gaebel, Gair, Gamba, Gammaitoni, Gaonkar, García-Quirós,
  Garufi, Gateley, Gaudio, Gaur, Gayathri, Gemme, Genin, Gennai, George,
  George, George, Gergely, Ghonge, Ghosh, Ghosh, Ghosh, Giacomazzo, Giaime,
  Giardina, Gibson, Gill, Glover, Gniesmer, Godwin, Goetz, Goetz, Goncharov,
  González, Castro, Gopakumar, Gossan, Gosselin, Gouaty, Grace, Grado,
  Granata, Grant, Gras, Grassia, Gray, Gray, Greco, Green, Green, Gretarsson,
  Grimaldi, Grimm, Groot, Grote, Grunewald, Gruning, Guidi, Gulati, Guo, Gupta,
  Gupta, Gupta, Gustafson, Gustafson, Haegel, Halim, Hall, Hall, Hamilton,
  Hammond, Haney, Hanke, Hanks, Hanna, Hannam, Hannuksela, Hansen, Hanson,
  Harder, Hardwick, Haris, Harms, Harry, Harry, Hasskew, Haster, Haughian,
  Hayes, Healy, Heidmann, Heintze, Heitmann, Hellman, Hello, Hemming, Hendry,
  Heng, Hennig, Heurs, Hild, Hinderer, Ho, Hochheim, Hofman, Holgado, Holland,
  Holt, Holz, Hopkins, Horst, Hough, Howell, Hoy, Huang, Hübner, Huerta, Huet,
  Hughey, Hui, Husa, Huttner, Huynh-Dinh, Idzkowski, Iess, Inchauspe, Ingram,
  Inta, Intini, Irwin, Isa, Isac, Isi, Iyer, Jacqmin, Jadhav, Jani, Janthalur,
  Jaranowski, Jariwala, Jenkins, Jiang, Johnson, Johnson-McDaniel, Jones,
  Jones, Jones, Jones, Jonker, Ju, Junker, Kalaghatgi, Kalogera, Kamai,
  Kandhasamy, Kang, Kanner, Kapadia, Karki, Kashyap, Kasprzack, Kastaun,
  Katsanevas, Katsavounidis, Katzman, Kaufer, Kawabe, Keerthana, Kéfélian,
  Keitel, Kennedy, Key, Khalili, Khan, Khan, Khazanov, Khetan, Khursheed,
  Kijbunchoo, Kim, Kim, Kim, Kim, Kim, Kim, Kimball, King, Kinley-Hanlon,
  Kirchhoff, Kissel, Kleybolte, Klika, Klimenko, Knowles, Koch, Koehlenbeck,
  Koekoek, Koley, Kondrashov, Kontos, Koper, Korobko, Korth, Kovalam, Kozak,
  Krämer, Kringel, Krishnendu, Królak, Krupinski, Kuehn, Kumar, Kumar, Kumar,
  Kumar, Kuo, Kutynia, Kwang, Lackey, Laghi, Lai, Lam, Landry, Landry, Lane,
  Lang, Lange, Lantz, Lanza, Lartaux-Vollard, Lasky, Laxen, Lazzarini, Lazzaro,
  Leaci, Leavey, Lecoeuche, Lee, Lee, Lee, Lee, Lee, Lee, Lehmann, Lenon,
  Leroy, Letendre, Levin, Li, Li, Li, Li, Li, Lin, Linde, Linker, Littenberg,
  Liu, Liu, Llorens-Monteagudo, Lo, London, Longo, Lorenzini, Loriette,
  Lormand, Losurdo, Lough, Lousto, Lovelace, Lower, Lucaccioni, Lück, Lumaca,
  Lundgren, Lynch, Ma, Macas, Macfoy, MacInnis, Macleod, Macquet, Hernandez,
  Magaña-Sandoval, Magee, Majorana, Maksimovic, Malik, Man, Mandic, Mangano,
  Mansell, Manske, Mantovani, Mapelli, Marchesoni, Marion, Márka, Márka,
  Markakis, Markosyan, Markowitz, Maros, Marquina, Marsat, Martelli, Martin,
  Martin, Martinez, Martynov, Masalehdan, Mason, Massera, Masserot, Massinger,
  Masso-Reid, Mastrogiovanni, Matas, Matichard, Matone, Mavalvala, McCann,
  McCarthy, McClelland, McCormick, McCuller, McGuire, McIsaac, McIver, McManus,
  McRae, McWilliams, Meacher, Meadors, Mehmet, Mehta, Meidam, Villa, Melatos,
  Mendell, Mercer, Mereni, Merfeld, Merilh, Merzougui, Meshkov, Messenger,
  Messick, Messina, Metzdorff, Meyers, Meylahn, Miani, Miao, Michel, Middleton,
  Milano, Miller, Millhouse, Mills, Milovich-Goff, Minazzoli, Minenkov,
  Mishkin, Mishra, Mistry, Mitra, Mitrofanov, Mitselmakher, Mittleman, Mo,
  Moffa, Mogushi, Mohapatra, Molina-Ruiz, Mondin, Montani, Moore, Moraru,
  Morawski, Moreno, Morisaki, Mours, Mow-Lowry, Muciaccia, Mukherjee,
  Mukherjee, Mukherjee, Mukherjee, Mukund, Mullavey, Munch, Muñiz, Muratore,
  Murray, Nagar, Nardecchia, Naticchioni, Nayak, Neil, Neilson, Nelemans,
  Nelson, Nery, Neunzert, Nevin, Ng, Ng, Nguyen, Nguyen, Nichols, Nichols,
  Nissanke, Nocera, North, Nuttall, Obergaulinger, Oberling, O’Brien,
  Oganesyan, Ogin, Oh, Oh, Ohme, Ohta, Okada, Oliver, Oppermann, Oram,
  O’Reilly, Ormiston, Ortega, O’Shaughnessy, Ossokine, Ottaway, Overmier,
  Owen, Pace, Pagano, Page, Pagliaroli, Pai, Pai, Palamos, Palashov, Palomba,
  Pan, Panda, Pang, Pankow, Pannarale, Pant, Paoletti, Paoli, Parida, Parker,
  Pascucci, Pasqualetti, Passaquieti, Passuello, Patil, Patricelli, Payne,
  Pearlstone, Pechsiri, Pedersen, Pedraza, Pedurand, Pele, Penn, Perego, Perez,
  Périgois, Perreca, Petermann, Pfeiffer, Phelps, Phukon, Piccinni, Pichot,
  Piergiovanni, Pierro, Pillant, Pinard, Pinto, Pirello, Pitkin, Plastino,
  Poggiani, Pong, Ponrathnam, Popolizio, Porter, Powell, Prajapati, Prasad,
  Prasai, Prasanna, Pratten, Prestegard, Principe, Prodi, Prokhorov, Punturo,
  Puppo, Pürrer, Qi, Quetschke, Quinonez, Raab, Raaijmakers, Radkins,
  Radulesco, Raffai, Raja, Rajan, Rajbhandari, Rakhmanov, Ramirez,
  Ramos-Buades, Rana, Rao, Rapagnani, Raymond, Razzano, Read, Regimbau, Rei,
  Reid, Reitze, Rettegno, Ricci, Richardson, Richardson, Ricker,
  Riemenschneider, Riles, Rizzo, Robertson, Robinet, Rocchi, Rolland, Rollins,
  Roma, Romanelli, Romano, Romel, Romie, Rose, Rose, Rose, Rosell, Rosińska,
  Rosofsky, Ross, Rowan, Roy, Rüdiger, Ruggi, Rutins, Ryan, Sachdev, Sadecki,
  Sakellariadou, Salafia, Salconi, Saleem, Samajdar, Sammut, Sanchez, Sanchez,
  Sanchis-Gual, Sanders, Santiago, Santos, Sarin, Sassolas, Sathyaprakash,
  Sauter, Savage, Schale, Scheel, Scheuer, Schmidt, Schnabel, Schofield,
  Schönbeck, Schreiber, Schulte, Schutz, Scott, Scott, Seidel, Sellers,
  Sengupta, Sennett, Sentenac, Sequino, Sergeev, Setyawati, Shaddock, Shaffer,
  Shahriar, Shaner, Sharma, Sharma, Shawhan, Shen, Shink, Shoemaker, Shoemaker,
  Shukla, ShyamSundar, Siellez, Sieniawska, Sigg, Singer, Singh, Singh,
  Singhal, Sintes, Sitmukhambetov, Skliris, Slagmolen, Slaven-Blair, Smith,
  Smith, Somala, Son, Soni, Sorazu, Sorrentino, Souradeep, Sowell, Spencer,
  Spera, Srivastava, Srivastava, Staats, Stachie, Standke, Steer, Steinke,
  Steinlechner, Steinlechner, Steinmeyer, Stevenson, Stocks, Stone, Stops,
  Strain, Stratta, Strigin, Strunk, Sturani, Stuver, Sudhir, Summerscales, Sun,
  Sunil, Sur, Suresh, Sutton, Swinkels, Szczepańczyk, Tacca, Tait, Talbot,
  Tanner, Tao, Tápai, Tapia, Tasson, Taylor, Tenorio, Terkowski, Thomas,
  Thomas, Thondapu, Thorne, Thrane, Tiwari, Tiwari, Tiwari, Toland, Tonelli,
  Tornasi, Torres-Forné, Torrie, Töyrä, Travasso, Traylor, Tringali,
  Tripathee, Trovato, Trozzo, Tsang, Tse, Tso, Tsukada, Tsuna, Tsutsui,
  Tuyenbayev, Ueno, Ugolini, Unnikrishnan, Urban, Usman, Vahlbruch, Vajente,
  Valdes, Valentini, Bakel, Beuzekom, Brand, Broeck, Vander-Hyde, Schaaf,
  VanHeijningen, Veggel, Vardaro, Varma, Vass, Vasúth, Vecchio, Vedovato,
  Veitch, Veitch, Venkateswara, Venugopalan, Verkindt, Vetrano, Viceré, Viets,
  Vinciguerra, Vine, Vinet, Vitale, Vo, Vocca, Vorvick, Vyatchanin, Wade, Wade,
  Wade, Walet, Walker, Wallace, Walsh, Wang, Wang, Wang, Wang, Ward, Warden,
  Warner, Was, Watchi, Weaver, Wei, Weinert, Weinstein, Weiss, Wellmann, Wen,
  Wessel, Weßels, Westhouse, Wette, Whelan, White, Whiting, Whittle, Wilken,
  Williams, Williamson, Willis, Willke, Winkler, Wipf, Wittel, Woan, Woehler,
  Wofford, Wright, Wu, Wysocki, Xiao, Xu, Yamamoto, Yancey, Yang, Yang, Yang,
  Yap, Yazback, Yeeles, Yu, Yu, Yuen, Zadrożny, Zadrożny, Zanolin, Zelenova,
  Zendri, Zevin, Zhang, Zhang, Zhang, Zhao, Zhao, Zhou, Zhou, Zhu, Zimmerman,
  Zucker, \& Zweizig}]{Abbott_2020}
Abbott, B.~P., Abbott, R., Abbott, T.~D., {et~al.} 2020, The Astrophysical
  Journal Letters, 892, L3

\bibitem[{Abbott {et~al.}(2019)Abbott, Abbott, Abbott, Abraham, Acernese,
  Ackley, Adams, Adhikari, Adya, Affeldt, Agathos, Agatsuma, Aggarwal, Aguiar,
  Aiello, Ain, Ajith, Allen, Allocca, Aloy, Altin, Amato, Ananyeva, Anderson,
  Anderson, Angelova, Antier, Appert, Arai, Araya, Areeda, Ar\`ene, Arnaud,
  Arun, Ascenzi, Ashton, Aston, Astone, Aubin, Aufmuth, AultONeal, Austin,
  Avendano, Avila-Alvarez, Babak, Bacon, Badaracco, Bader, Bae, Baker,
  Baldaccini, Ballardin, Ballmer, Banagiri, Barayoga, Barclay, Barish, Barker,
  Barkett, Barnum, Barone, Barr, Barsotti, Barsuglia, Barta, Bartlett, Bartos,
  Bassiri, Basti, Bawaj, Bayley, Bazzan, B\'ecsy, Bejger, Belahcene, Bell,
  Beniwal, Berger, Bergmann, Bernuzzi, Bero, Berry, Bersanetti, Bertolini,
  Betzwieser, Bhandare, Bidler, Bilenko, Bilgili, Billingsley, Birch, Birney,
  Birnholtz, Biscans, Biscoveanu, Bisht, Bitossi, Bizouard, Blackburn,
  Blackman, Blair, Blair, Blair, Bloemen, Bode, Boer, Boetzel, Bogaert, Bondu,
  Bonilla, Bonnand, Booker, Boom, Booth, Bork, Boschi, Bose, Bossie, Bossilkov,
  Bosveld, Bouffanais, Bozzi, Bradaschia, Brady, Bramley, Branchesi, Brau,
  Briant, Briggs, Brighenti, Brillet, Brinkmann, Brisson, Brockill, Brooks,
  Brown, Brunett, Buikema, Bulik, Bulten, Buonanno, Buskulic,
  Bustamante~Rosell, Buy, Byer, Cabero, Cadonati, Cagnoli, Cahillane,
  Calder\'on~Bustillo, Callister, Calloni, Camp, Campbell, Canepa, Cannon, Cao,
  Cao, Capocasa, Carbognani, Caride, Carney, Carullo, Casanueva~Diaz,
  Casentini, Caudill, Cavagli\`a, Cavalier, Cavalieri, Cella, Cerd\'a-Dur\'an,
  Cerretani, Cesarini, Chaibi, Chakravarti, Chamberlin, Chan, Chao, Charlton,
  Chase, Chassande-Mottin, Chatterjee, Chaturvedi, Chatziioannou, Cheeseboro,
  Chen, Chen, Chen, Cheng, Cheong, Chia, Chincarini, Chiummo, Cho, Cho, Cho,
  Christensen, Chu, Chua, Chung, Chung, Ciani, Ciobanu, Ciolfi, Cipriano,
  Cirone, Clara, Clark, Clearwater, Cleva, Cocchieri, Coccia, Cohadon, Cohen,
  Colgan, Colleoni, Collette, Collins, Cominsky, Constancio, Conti, Cooper,
  Corban, Corbitt, Cordero-Carri\'on, Corley, Cornish, Corsi, Cortese, Costa,
  Cotesta, Coughlin, Coughlin, Coulon, Countryman, Couvares, Covas, Cowan,
  Coward, Cowart, Coyne, Coyne, Creighton, Creighton, Cripe, Croquette,
  Crowder, Cullen, Cumming, Cunningham, Cuoco, Canton, D\'alya, Danilishin,
  D'Antonio, Danzmann, Dasgupta, Da~Silva~Costa, Datrier, Dattilo, Dave,
  Davier, Davis, Daw, DeBra, Deenadayalan, Degallaix, De~Laurentis,
  Del\'eglise, Del~Pozzo, DeMarchi, Demos, Dent, De~Pietri, Derby, De~Rosa,
  De~Rossi, DeSalvo, de~Varona, Dhurandhar, D\'{\i}az, Dietrich, Di~Fiore,
  Di~Giovanni, Di~Girolamo, Di~Lieto, Ding, Di~Pace, Di~Palma, Di~Renzo,
  Dmitriev, Doctor, Donovan, Dooley, Doravari, Dorrington, Downes, Drago,
  Driggers, Du, Ducoin, Dupej, Dwyer, Easter, Edo, Edwards, Effler, Ehrens,
  Eichholz, Eikenberry, Eisenmann, Eisenstein, Essick, Estelles, Estevez,
  Etienne, Etzel, Evans, Evans, Fafone, Fair, Fairhurst, Fan, Farinon, Farr,
  Farr, Fauchon-Jones, Favata, Fays, Fazio, Fee, Feicht, Fejer, Feng,
  Fernandez-Galiana, Ferrante, Ferreira, Ferreira, Ferrini, Fidecaro, Fiori,
  Fiorucci, Fishbach, Fisher, Fishner, Fitz-Axen, Flaminio, Fletcher, Flynn,
  Fong, Font, Forsyth, Fournier, Frasca, Frasconi, Frei, Freise, Frey, Frey,
  Fritschel, Frolov, Fulda, Fyffe, Gabbard, Gadre, Gaebel, Gair, Gammaitoni,
  Ganija, Gaonkar, Garcia, Garc\'{\i}a-Quir\'os, Garufi, Gateley, Gaudio, Gaur,
  Gayathri, Gemme, Genin, Gennai, George, George, Gergely, Germain, Ghonge,
  Ghosh, Ghosh, Ghosh, Giacomazzo, Giaime, Giardina, Giazotto, Gill, Giordano,
  Glover, Godwin, Goetz, Goetz, Goncharov, Gonz\'alez, Gonzalez~Castro,
  Gopakumar, Gorodetsky, Gossan, Gosselin, Gouaty, Grado, Graef, Granata,
  Grant, Gras, Grassia, Gray, Gray, Greco, Green, Green, Gretarsson, Groot,
  Grote, Grunewald, Gruning, Guidi, Gulati, Guo, Gupta, Gupta, Gustafson,
  Gustafson, Haegel, Halim, Hall, Hall, Hamilton, Hammond, Haney, Hanke, Hanks,
  Hanna, Hannam, Hannuksela, Hanson, Hardwick, Haris, Harms, Harry, Harry,
  Haster, Haughian, Hayes, Healy, Heidmann, Heintze, Heitmann, Hello, Hemming,
  Hendry, Heng, Hennig, Heptonstall, Hernandez~Vivanco, Heurs, Hild, Hinderer,
  Hoak, Hochheim, Hofman, Holgado, Holland, Holt, Holz, Hopkins, Horst, Hough,
  Howell, Hoy, Hreibi, Huang, Huerta, Huet, Hughey, Hulko, Husa, Huttner,
  Huynh-Dinh, Idzkowski, Iess, Ingram, Inta, Intini, Irwin, Isa, Isac, Isi,
  Iyer, Izumi, Jacqmin, Jadhav, Jani, Janthalur, Jaranowski, Jenkins, Jiang,
  Johnson, Johnson-McDaniel, Jones, Jones, Jones, Jonker, Ju, Junker,
  Kalaghatgi, Kalogera, Kamai, Kandhasamy, Kang, Kanner, Kapadia, Karki,
  Karvinen, Kashyap, Kasprzack, Katsanevas, Katsavounidis, Katzman, Kaufer,
  Kawabe, Keerthana, K\'ef\'elian, Keitel, Kennedy, Key, Khalili, Khan, Khan,
  Khan, Khan, Khazanov, Khursheed, Kijbunchoo, Kim, Kim, Kim, Kim, Kim, Kim,
  Kimball, King, King, Kinley-Hanlon, Kirchhoff, Kissel, Kleybolte, Klika,
  Klimenko, Knowles, Koch, Koehlenbeck, Koekoek, Koley, Kondrashov, Kontos,
  Koper, Korobko, Korth, Kowalska, Kozak, Kringel, Krishnendu, Kr\'olak, Kuehn,
  Kumar, Kumar, Kumar, Kumar, Kuo, Kutynia, Kwang, Lackey, Lai, Lam, Landry,
  Lane, Lang, Lange, Lantz, Lanza, Lartaux-Vollard, Lasky, Laxen, Lazzarini,
  Lazzaro, Leaci, Leavey, Lecoeuche, Lee, Lee, Lee, Lee, Lee, Lee, Lehmann,
  Lenon, Leroy, Letendre, Levin, Li, Li, Li, Li, Lin, Linde, Linker,
  Littenberg, Liu, Liu, Lo, Lockerbie, London, Longo, Lorenzini, Loriette,
  Lormand, Losurdo, Lough, Lousto, Lovelace, Lower, L\"uck, Lumaca, Lundgren,
  Lynch, Ma, Macas, Macfoy, MacInnis, Macleod, Macquet, Maga\~na Sandoval,
  Maga\~na Zertuche, Magee, Majorana, Maksimovic, Malik, Man, Mandic, Mangano,
  Mansell, Manske, Mantovani, Marchesoni, Marion, M\'arka, M\'arka, Markakis,
  Markosyan, Markowitz, Maros, Marquina, Marsat, Martelli, Martin, Martin,
  Martynov, Mason, Massera, Masserot, Massinger, Masso-Reid, Mastrogiovanni,
  Matas, Matichard, Matone, Mavalvala, Mazumder, McCann, McCarthy, McClelland,
  McCormick, McCuller, McGuire, McIver, McManus, McRae, McWilliams, Meacher,
  Meadors, Mehmet, Mehta, Meidam, Melatos, Mendell, Mercer, Mereni, Merilh,
  Merzougui, Meshkov, Messenger, Messick, Metzdorff, Meyers, Miao, Michel,
  Middleton, Mikhailov, Milano, Miller, Miller, Millhouse, Mills,
  Milovich-Goff, Minazzoli, Minenkov, Mishkin, Mishra, Mistry, Mitra,
  Mitrofanov, Mitselmakher, Mittleman, Mo, Moffa, Mogushi, Mohapatra, Montani,
  Moore, Moraru, Moreno, Morisaki, Mours, Mow-Lowry, Mukherjee, Mukherjee,
  Mukherjee, Mukund, Mullavey, Munch, Mu\~niz, Muratore, Murray, Nagar,
  Nardecchia, Naticchioni, Nayak, Neilson, Nelemans, Nelson, Nery, Neunzert,
  Ng, Ng, Nguyen, Nichols, Nielsen, Nissanke, Nitz, Nocera, North, Nuttall,
  Obergaulinger, Oberling, O'Brien, O'Dea, Ogin, Oh, Oh, Ohme, Ohta, Okada,
  Oliver, Oppermann, Oram, O'Reilly, Ormiston, Ortega, O'Shaughnessy, Ossokine,
  Ottaway, Overmier, Owen, Pace, Pagano, Page, Pai, Pai, Palamos, Palashov,
  Palomba, Pal-Singh, Pan, Pang, Pang, Pankow, Pannarale, Pant, Paoletti,
  Paoli, Papa, Parida, Parker, Pascucci, Pasqualetti, Passaquieti, Passuello,
  Patil, Patricelli, Pearlstone, Pedersen, Pedraza, Pedurand, Pele, Penn,
  Perego, Perez, Perreca, Pfeiffer, Phelps, Phukon, Piccinni, Pichot,
  Piergiovanni, Pillant, Pinard, Pirello, Pitkin, Poggiani, Pong, Ponrathnam,
  Popolizio, Porter, Powell, Prajapati, Prasad, Prasai, Prasanna, Pratten,
  Prestegard, Privitera, Prodi, Prokhorov, Puncken, Punturo, Puppo, P\"urrer,
  Qi, Quetschke, Quinonez, Quintero, Quitzow-James, Raab, Radkins, Radulescu,
  Raffai, Raja, Rajan, Rajbhandari, Rakhmanov, Ramirez, Ramos-Buades, Rana,
  Rao, Rapagnani, Raymond, Razzano, Read, Regimbau, Rei, Reid, Reitze, Ren,
  Ricci, Richardson, Richardson, Ricker, Riemenschneider, Riles, Rizzo,
  Robertson, Robie, Robinet, Rocchi, Rolland, Rollins, Roma, Romanelli, Romano,
  Romel, Romie, Rose, Rosi\ifmmode~\acute{n}\else \'{n}\fi{}ska, Rosofsky,
  Ross, Rowan, R\"udiger, Ruggi, Rutins, Ryan, Sachdev, Sadecki, Sakellariadou,
  Salafia, Salconi, Saleem, Salemi, Samajdar, Sammut, Sanchez, Sanchez,
  Sanchis-Gual, Sandberg, Sanders, Santiago, Sarin, Sassolas, Sathyaprakash,
  Saulson, Sauter, Savage, Schale, Scheel, Scheuer, Schmidt, Schnabel,
  Schofield, Sch\"onbeck, Schreiber, Schulte, Schutz, Schwalbe, Scott, Scott,
  Seidel, Sellers, Sengupta, Sennett, Sentenac, Sequino, Sergeev, Setyawati,
  Shaddock, Shaffer, Shahriar, Shaner, Shao, Sharma, Shawhan, Shen, Shink,
  Shoemaker, Shoemaker, ShyamSundar, Siellez, Sieniawska, Sigg, Silva, Singer,
  Singh, Singhal, Sintes, Sitmukhambetov, Skliris, Slagmolen, Slaven-Blair,
  Smith, Smith, Somala, Son, Sorazu, Sorrentino, Souradeep, Sowell, Spencer,
  Srivastava, Srivastava, Staats, Stachie, Standke, Steer, Steinke,
  Steinlechner, Steinlechner, Steinmeyer, Stevenson, Stocks, Stone, Stops,
  Strain, Stratta, Strigin, Strunk, Sturani, Stuver, Sudhir, Summerscales, Sun,
  Sunil, Suresh, Sutton, Swinkels, Szczepa\ifmmode~\acute{n}\else
  \'{n}\fi{}czyk, Tacca, Tait, Talbot, Talukder, Tanner, T\'apai, Taracchini,
  Tasson, Taylor, Thies, Thomas, Thomas, Thondapu, Thorne, Thrane, Tiwari,
  Tiwari, Tiwari, Toland, Tonelli, Tornasi, Torres-Forn\'e, Torrie, T\"oyr\"a,
  Travasso, Traylor, Tringali, Trovato, Trozzo, Trudeau, Tsang, Tse, Tso,
  Tsukada, Tsuna, Tuyenbayev, Ueno, Ugolini, Unnikrishnan, Urban, Usman,
  Vahlbruch, Vajente, Valdes, van Bakel, van Beuzekom, van~den Brand, Van
  Den~Broeck, Vander-Hyde, van Heijningen, van~der Schaaf, van Veggel, Vardaro,
  Varma, Vass, Vas\'uth, Vecchio, Vedovato, Veitch, Veitch, Venkateswara,
  Venugopalan, Verkindt, Vetrano, Vicer\'e, Viets, Vine, Vinet, Vitale, Vo,
  Vocca, Vorvick, Vyatchanin, Wade, Wade, Wade, Walet, Walker, Wallace, Walsh,
  Wang, Wang, Wang, Wang, Wang, Ward, Warden, Warner, Was, Watchi, Weaver, Wei,
  Weinert, Weinstein, Weiss, Wellmann, Wen, Wessel, We\ss{}els, Westhouse,
  Wette, Whelan, White, Whiting, Whittle, Wilken, Williams, Williamson, Willis,
  Willke, Wimmer, Winkler, Wipf, Wittel, Woan, Woehler, Wofford, Worden,
  Wright, Wu, Wysocki, Xiao, Yamamoto, Yancey, Yang, Yap, Yazback, Yeeles, Yu,
  Yu, Yuen, Yvert, Zadro\ifmmode~\dot{z}\else \.{z}\fi{}ny, Zanolin, Zappa,
  Zelenova, Zendri, Zevin, Zhang, Zhang, Zhang, Zhao, Zhou, Zhou, Zhu,
  Zimmerman, Zlochower, Zucker, \& Zweizig}]{abbottPhysRevX.9.031040}
Abbott, B.~P., Abbott, R., Abbott, T.~D., {et~al.} 2019, Phys. Rev. X, 9,
  031040

\bibitem[{Abbott {et~al.}(2021)Abbott, Abbott, Abraham, Acernese, Ackley,
  Adams, Adams, Adhikari, Adya, Affeldt, Agathos, Agatsuma, Aggarwal, Aguiar,
  Aiello, Ain, Ajith, Akcay, Allen, Allocca, Altin, Amato, Anand, Ananyeva,
  Anderson, Anderson, Angelova, Ansoldi, Antelis, Antier, Appert, Arai, Araya,
  Areeda, Ar\`ene, Arnaud, Aronson, Arun, Asali, Ascenzi, Ashton, Aston,
  Astone, Aubin, Aufmuth, AultONeal, Austin, Avendano, Babak, Badaracco, Bader,
  Bae, Baer, Bagnasco, Baird, Ball, Ballardin, Ballmer, Bals, Balsamo, Baltus,
  Banagiri, Bankar, Bankar, Barayoga, Barbieri, Barish, Barker, Barneo, Barnum,
  Barone, Barr, Barsotti, Barsuglia, Barta, Bartlett, Bartos, Bassiri, Basti,
  Bawaj, Bayley, Bazzan, Becher, B\'ecsy, Bedakihale, Bejger, Belahcene,
  Beniwal, Benjamin, Bennett, Bentley, Bergamin, Berger, Bergmann, Bernuzzi,
  Berry, Bersanetti, Bertolini, Betzwieser, Bhandare, Bhandari, Bhattacharjee,
  Bidler, Bilenko, Billingsley, Birney, Birnholtz, Biscans, Bischi, Biscoveanu,
  Bisht, Bitossi, Bizouard, Blackburn, Blackman, Blair, Blair, Blair, Blanch,
  Bobba, Bode, Boer, Boetzel, Bogaert, Boldrini, Bondu, Bonilla, Bonnand,
  Booker, Boom, Bork, Boschi, Bose, Bossilkov, Boudart, Bouffanais, Bozzi,
  Bradaschia, Brady, Bramley, Branchesi, Brau, Breschi, Briant, Briggs,
  Brighenti, Brillet, Brinkmann, Brockill, Brooks, Brooks, Brown, Brunett,
  Bruno, Bruntz, Buikema, Bulik, Bulten, Buonanno, Buscicchio, Buskulic, Byer,
  Cabero, Cadonati, Caesar, Cagnoli, Cahillane, Calder\'on~Bustillo, Callaghan,
  Callister, Calloni, Camp, Canepa, Cannon, Cao, Cao, Carapella, Carbognani,
  Carney, Carpinelli, Carullo, Carver, Casanueva~Diaz, Casentini, Caudill,
  Cavagli\`a, Cavalier, Cavalieri, Cella, Cerd\'a-Dur\'an, Cesarini, Chaibi,
  Chakravarti, Chan, Chan, Chandra, Chanial, Chao, Charlton, Chase,
  Chassande-Mottin, Chatterjee, Chattopadhyay, Chaturvedi, Chatziioannou, Chen,
  Chen, Chen, Chen, Cheng, Cheong, Chia, Chiadini, Chierici, Chincarini,
  Chiummo, Cho, Cho, Cho, Choate, Christensen, Chu, Chua, Chung, Chung, Ciani,
  Ciecielag, Cie\ifmmode~\acute{s}\else \'{s}\fi{}lar, Cifaldi, Ciobanu,
  Ciolfi, Cipriano, Cirone, Clara, Clark, Clark, Clarke, Clearwater, Clesse,
  Cleva, Coccia, Cohadon, Cohen, Colleoni, Collette, Collins, Colpi,
  Constancio, Conti, Cooper, Corban, Corbitt, Cordero-Carri\'on, Corezzi,
  Corley, Cornish, Corre, Corsi, Cortese, Costa, Cotesta, Coughlin, Coughlin,
  Coulon, Countryman, Cousins, Couvares, Covas, Coward, Cowart, Coyne, Coyne,
  Creighton, Creighton, Croquette, Crowder, Cudell, Cullen, Cumming, Cummings,
  Cunningham, Cuoco, Cury\l{}o, Canton, D\'alya, Dana, DaneshgaranBajastani,
  D'Angelo, Danila, Danilishin, D'Antonio, Danzmann, Darsow-Fromm, Dasgupta,
  Datrier, Dattilo, Dave, Davier, Davies, Davis, Daw, Dean, DeBra,
  Deenadayalan, Degallaix, De~Laurentis, Del\'eglise, Del~Favero, De~Lillo,
  De~Lillo, Del~Pozzo, DeMarchi, De~Matteis, D'Emilio, Demos, Denker, Dent,
  Depasse, De~Pietri, De~Rosa, De~Rossi, DeSalvo, de~Varona, Dhurandhar,
  D\'{\i}az, Diaz-Ortiz, Didio, Dietrich, Di~Fiore, DiFronzo, Di~Giorgio,
  Di~Giovanni, Di~Giovanni, Di~Girolamo, Di~Lieto, Ding, Di~Pace, Di~Palma,
  Di~Renzo, Divakarla, Dmitriev, Doctor, D'Onofrio, Donovan, Dooley, Doravari,
  Dorrington, Downes, Drago, Driggers, Du, Ducoin, Dupej, Durante, D'Urso,
  Duverne, Dwyer, Easter, Eddolls, Edelman, Edo, Edy, Effler, Eichholz,
  Eikenberry, Eisenmann, Eisenstein, Ejlli, Errico, Essick, Estell\'es,
  Estevez, Etienne, Etzel, Evans, Evans, Ewing, Fafone, Fair, Fairhurst, Fan,
  Farah, Farinon, Farr, Farr, Fauchon-Jones, Favata, Fays, Fazio, Feicht,
  Fejer, Feng, Fenyvesi, Ferguson, Fernandez-Galiana, Ferrante, Ferreira,
  Fidecaro, Figura, Fiori, Fiorucci, Fishbach, Fisher, Fishner, Fittipaldi,
  Fitz-Axen, Fiumara, Flaminio, Floden, Flynn, Fong, Font, Forsyth, Fournier,
  Frasca, Frasconi, Frei, Freise, Frey, Frey, Fritschel, Frolov, Fronz\'e,
  Fulda, Fyffe, Gabbard, Gadre, Gaebel, Gair, Gais, Galaudage, Gamba,
  Ganapathy, Ganguly, Gaonkar, Garaventa, Garc\'{\i}a-Quir\'os, Garufi,
  Gateley, Gaudio, Gayathri, Gemme, Gennai, George, George, George, Gergely,
  Ghonge, Ghosh, Ghosh, Ghosh, Giacomazzo, Giacoppo, Giaime, Giardina, Gibson,
  Gier, Gill, Giri, Glanzer, Gleckl, Godwin, Goetz, Goetz, Gohlke, Goncharov,
  Gonz\'alez, Gopakumar, Gossan, Gosselin, Gouaty, Grace, Grado, Granata,
  Granata, Grant, Gras, Grassia, Gray, Gray, Greco, Green, Green, Gretarsson,
  Griggs, Grignani, Grimaldi, Grimes, Grimm, Grote, Grunewald, Gruning,
  Guerrero, Guidi, Guimaraes, Guix\'e, Gulati, Guo, Gupta, Gupta, Gupta,
  Gustafson, Gustafson, Guzman, Haegel, Halim, Hall, Hamilton, Hammond, Haney,
  Hanke, Hanks, Hanna, Hannam, Hannuksela, Hannuksela, Hansen, Hansen, Hanson,
  Harder, Hardwick, Haris, Harms, Harry, Harry, Hartwig, Hasskew, Haster,
  Haughian, Hayes, Healy, Heidmann, Heintze, Heinze, Heinzel, Heitmann,
  Hellman, Hello, Helmling-Cornell, Hemming, Hendry, Heng, Hennes, Hennig,
  Hennig, Hernandez~Vivanco, Heurs, Hild, Hill, Hines, Hochheim, Hofgard,
  Hofman, Hohmann, Holgado, Holland, Hollows, Holmes, Holt, Holz, Hopkins,
  Horst, Hough, Howell, Hoy, Hoyland, Huang, H\"ubner, Huddart, Huerta, Hughey,
  Hui, Husa, Huttner, Hutzler, Huxford, Huynh-Dinh, Idzkowski, Iess, Imperato,
  Inchauspe, Ingram, Intini, Isi, Iyer, JaberianHamedan, Jacqmin, Jadhav,
  Jadhav, James, Jani, Janssens, Janthalur, Jaranowski, Jariwala, Jaume,
  Jenkins, Jeunon, Jiang, Johns, Johnson-McDaniel, Jones, Jones, Jones, Jones,
  Jones, Jonker, Ju, Junker, Kalaghatgi, Kalogera, Kamai, Kandhasamy, Kang,
  Kanner, Kapadia, Kapasi, Karathanasis, Karki, Kashyap, Kasprzack, Kastaun,
  Katsanevas, Katsavounidis, Katzman, Kawabe, K\'ef\'elian, Keitel, Key,
  Khadka, Khalili, Khan, Khan, Khazanov, Khetan, Khursheed, Kijbunchoo, Kim,
  Kim, Kim, Kim, Kim, Kim, Kimball, King, Kinley-Hanlon, Kirchhoff, Kissel,
  Kleybolte, Klimenko, Knowles, Knyazev, Koch, Koehlenbeck, Koekoek, Koley,
  Kolstein, Komori, Kondrashov, Kontos, Koper, Korobko, Korth, Kovalam, Kozak,
  Kr\"amer, Kringel, Krishnendu, Kr\'olak, Kuehn, Kumar, Kumar, Kumar, Kumar,
  Kuns, Kwang, Lackey, Laghi, Lalande, Lam, Lamberts, Landry, Lane, Lang,
  Lange, Lantz, Lanza, La~Rosa, Lartaux-Vollard, Lasky, Laxen, Lazzarini,
  Lazzaro, Leaci, Leavey, Lecoeuche, Lee, Lee, Lee, Lee, Lehmann, Leon, Leroy,
  Letendre, Levin, Li, Li, Li, Li, Li, Linde, Linker, Linley, Littenberg, Liu,
  Liu, Llorens-Monteagudo, Lo, Lockwood, London, Longo, Lorenzini, Loriette,
  Lormand, Losurdo, Lough, Lousto, Lovelace, L\"uck, Lumaca, Lundgren, Ma,
  Macas, MacInnis, Macleod, MacMillan, Macquet, Maga\~na Hernandez, Maga\~na
  Sandoval, Magazz\`u, Magee, Majorana, Maksimovic, Maliakal, Malik, Man,
  Mandic, Mangano, Mansell, Manske, Mantovani, Mapelli, Marchesoni, Marion,
  M\'arka, M\'arka, Markakis, Markosyan, Markowitz, Maros, Marquina, Marsat,
  Martelli, Martin, Martin, Martinez, Martinez, Martynov, Masalehdan, Mason,
  Massera, Masserot, Massinger, Masso-Reid, Mastrogiovanni, Matas,
  Mateu-Lucena, Matichard, Matiushechkina, Mavalvala, Maynard, McCann,
  McCarthy, McClelland, McCormick, McCuller, McGuire, McIsaac, McIver, McManus,
  McRae, McWilliams, Meacher, Meadors, Mehmet, Mehta, Melatos, Melchor,
  Mendell, Menendez-Vazquez, Mercer, Mereni, Merfeld, Merilh, Merritt,
  Merzougui, Meshkov, Messenger, Messick, Metzdorff, Meyers, Meylahn, Mhaske,
  Miani, Miao, Michaloliakos, Michel, Middleton, Milano, Miller, Millhouse,
  Mills, Milotti, Milovich-Goff, Minazzoli, Minenkov, Mir, Mishkin, Mishra,
  Mistry, Mitra, Mitrofanov, Mitselmakher, Mittleman, Mo, Mogushi, Mohapatra,
  Mohite, Molina, Molina-Ruiz, Mondin, Montani, Moore, Moraru, Morawski,
  Moreno, Morisaki, Mours, Mow-Lowry, Mozzon, Muciaccia, Mukherjee, Mukherjee,
  Mukherjee, Mukherjee, Mukund, Mullavey, Munch, Mu\~niz, Murray, Nadji, Nagar,
  Nardecchia, Naticchioni, Nayak, Neil, Neilson, Nelemans, Nelson, Nery,
  Neunzert, Nitz, Ng, Ng, Nguyen, Nguyen, Nguyen, Nichols, Nissanke, Nocera,
  Noh, North, Nothard, Nuttall, Oberling, O'Brien, O'Dell, Oganesyan, Ogin, Oh,
  Oh, Ohme, Ohta, Okada, Olivetto, Oppermann, Oram, O'Reilly, Ormiston, Ortega,
  O'Shaughnessy, Ossokine, Osthelder, Ottaway, Overmier, Owen, Pace, Pagano,
  Page, Pagliaroli, Pai, Pai, Palamos, Palashov, Palomba, Pan, Panda, Pang,
  Pankow, Pannarale, Pant, Paoletti, Paoli, Paolone, Parker, Pascucci,
  Pasqualetti, Passaquieti, Passuello, Patel, Patricelli, Payne, Pechsiri,
  Pedraza, Pegoraro, Pele, Penn, Perego, Perez, P\'erigois, Perreca, Perri\`es,
  Petermann, Petterson, Pfeiffer, Pham, Phukon, Piccinni, Pichot, Piendibene,
  Piergiovanni, Pierini, Pierro, Pillant, Pilo, Pinard, Pinto, Piotrzkowski,
  Pirello, Pitkin, Placidi, Plastino, Pluchar, Poggiani, Polini, Pong,
  Ponrathnam, Popolizio, Porter, Poverman, Powell, Pracchia, Prajapati, Prasai,
  Prasanna, Pratten, Prestegard, Principe, Prodi, Prokhorov, Prosposito,
  Prudenzi, Puecher, Punturo, Puosi, Puppo, P\"urrer, Qi, Quetschke, Quinonez,
  Quitzow-James, Raab, Raaijmakers, Radkins, Radulesco, Raffai, Rafferty, Rail,
  Raja, Rajan, Rajbhandari, Rakhmanov, Ramirez, Ramirez, Ramos-Buades, Rana,
  Rao, Rapagnani, Rapol, Ratto, Raymond, Razzano, Read, Regimbau, Rei, Reid,
  Reitze, Rettegno, Ricci, Richardson, Richardson, Richardson, Ricker,
  Riemenschneider, Riles, Rizzo, Robertson, Robinet, Rocchi, Rocha, Rodriguez,
  Rodriguez-Soto, Rolland, Rollins, Roma, Romanelli, Romano, Romel, Romero,
  Romero-Shaw, Romie, Ronchini, Rose, Rose, Rose, Rosell,
  Rosi\ifmmode~\acute{n}\else \'{n}\fi{}ska, Rosofsky, Ross, Rowan, Rowlinson,
  Roy, Roy, Ruggi, Ryan, Sachdev, Sadecki, Sadiq, Sakellariadou, Salafia,
  Salconi, Saleem, Samajdar, Sanchez, Sanchez, Sanchez, Sanchis-Gual, Sanders,
  Sandles, Santiago, Santos, Saravanan, Sarin, Sassolas, Sathyaprakash, Sauter,
  Savage, Savant, Sawant, Sayah, Schaetzl, Schale, Scheel, Scheuer,
  Schindler-Tyka, Schmidt, Schnabel, Schofield, Sch\"onbeck, Schreiber,
  Schulte, Schutz, Schwarm, Schwartz, Scott, Scott, Seglar-Arroyo, Seidel,
  Sellers, Sengupta, Sennett, Sentenac, Sequino, Sergeev, Setyawati, Shaffer,
  Shahriar, Sharifi, Sharma, Sharma, Shawhan, Shen, Shikauchi, Shink,
  Shoemaker, Shoemaker, Shukla, ShyamSundar, Sieniawska, Sigg, Singer, Singh,
  Singh, Singha, Singhal, Sintes, Sipala, Skliris, Slagmolen, Slaven-Blair,
  Smetana, Smith, Smith, Somala, Son, Soni, Soni, Sorazu, Sordini, Sorrentino,
  Sorrentino, Soulard, Souradeep, Sowell, Spencer, Spera, Srivastava,
  Srivastava, Staats, Stachie, Steer, Steinhoff, Steinke, Steinlechner,
  Steinlechner, Steinmeyer, Stevenson, Stolle-McAllister, Stops, Stover,
  Strain, Stratta, Strunk, Sturani, Stuver, S\"udbeck, Sudhagar, Sudhir, Suh,
  Summerscales, Sun, Sun, Sunil, Sur, Suresh, Sutton, Swinkels,
  Szczepa\ifmmode~\acute{n}\else \'{n}\fi{}czyk, Tacca, Tait, Talbot,
  Tanasijczuk, Tanner, Tao, Tapia, Tapia San~Martin, Tasson, Taylor, Tenorio,
  Terkowski, Thirugnanasambandam, Thomas, Thomas, Thomas, Thompson, Thondapu,
  Thorne, Thrane, Tiwari, Tiwari, Tiwari, Toland, Tolley, Tonelli, Tornasi,
  Torres-Forn\'e, Torrie, e~Melo, T\"oyr\"a, Tran, Trapananti, Travasso,
  Traylor, Tringali, Tripathee, Trovato, Trudeau, Tsai, Tsang, Tse, Tso,
  Tsukada, Tsuna, Tsutsui, Turconi, Ubhi, Udall, Ueno, Ugolini, Unnikrishnan,
  Urban, Usman, Utina, Vahlbruch, Vajente, Vajpeyi, Valdes, Valentini, Valsan,
  van Bakel, van Beuzekom, van~den Brand, Van Den~Broeck, Vander-Hyde, van~der
  Schaaf, van Heijningen, Vardaro, Vargas, Varma, Vass, Vas\'uth, Vecchio,
  Vedovato, Veitch, Veitch, Venkateswara, Venneberg, Venugopalan, Verkindt,
  Verma, Veske, Vetrano, Vicer\'e, Viets, Vijaykumar, Villa-Ortega, Vinet,
  Vitale, Vo, Vocca, Vorvick, Vyatchanin, Wade, Wade, Wade, Walet, Walker,
  Wallace, Wallace, Walsh, Wang, Wang, Wang, Wang, Ward, Warner, Was,
  Washington, Watchi, Weaver, Wei, Weinert, Weinstein, Weiss, Wellmann, Wen,
  We\ss{}els, Westhouse, Wette, Whelan, White, White, Whiting, Whittle, Wilken,
  Williams, Williams, Williamson, Willis, Willke, Wilson, Wimmer, Winkler,
  Wipf, Woan, Woehler, Wofford, Wong, Wrangel, Wright, Wu, Wysocki, Xiao,
  Yamamoto, Yang, Yang, Yang, Yap, Yeeles, Yoon, Yu, Yu, Yuen,
  Zadro\ifmmode~\dot{z}\else \.{z}\fi{}ny, Zanolin, Zelenova, Zendri, Zevin,
  Zhang, Zhang, Zhang, Zhang, Zhao, Zhao, Zheng, Zhou, Zhou, Zhu, Zimmerman,
  Zlochower, Zucker, \& Zweizig}]{abbottPhysRevX.11.021053}
Abbott, R., Abbott, T.~D., Abraham, S., {et~al.} 2021, Phys. Rev. X, 11, 021053

\bibitem[{{Agrawal} {et~al.}(2023){Agrawal}, {Hurley}, {Stevenson},
  {Rodriguez}, {Sz{\'e}csi}, \& {Kemp}}]{Agrawal2023}
{Agrawal}, P., {Hurley}, J., {Stevenson}, S., {et~al.} 2023, \mnras, 525, 933

\bibitem[{{Agrawal} {et~al.}(2020){Agrawal}, {Hurley}, {Stevenson},
  {Sz{\'e}csi}, \& {Flynn}}]{Agrawal2020}
{Agrawal}, P., {Hurley}, J., {Stevenson}, S., {Sz{\'e}csi}, D., \& {Flynn}, C.
  2020, \mnras, 497, 4549

\bibitem[{{Andrews} {et~al.}(2024){Andrews}, {Bavera}, {Briel}, {Chattaraj},
  {Dotter}, {Fragos}, {Gallegos-Garcia}, {Gossage}, {Kalogera}, {Kasdagli},
  {Katsaggelos}, {Kimball}, {Kovlakas}, {Kruckow}, {Liotine}, {Misra}, {Rocha},
  {Souropanis}, {Srivastava}, {Sun}, {Teng}, {Xing}, {Zapartas}, \&
  {Zevin}}]{Andrews2024arXiv}
{Andrews}, J.~J., {Bavera}, S.~S., {Briel}, M., {et~al.} 2024, arXiv e-prints,
  arXiv:2411.02376

\bibitem[{{Andronov} {et~al.}(2006){Andronov}, {Pinsonneault}, \&
  {Terndrup}}]{Andronov_2006}
{Andronov}, N., {Pinsonneault}, M.~H., \& {Terndrup}, D.~M. 2006, \apj, 646,
  1160

\bibitem[{{Banerjee} {et~al.}(2010){Banerjee}, {Baumgardt}, \&
  {Kroupa}}]{Banerjee2010MNRAS.402..371B}
{Banerjee}, S., {Baumgardt}, H., \& {Kroupa}, P. 2010, \mnras, 402, 371

\bibitem[{{Bavera} {et~al.}(2021){Bavera}, {Fragos}, {Zevin}, {Berry},
  {Marchant}, {Andrews}, {Coughlin}, {Dotter}, {Kovlakas}, {Misra},
  {Serra-Perez}, {Qin}, {Rocha}, {Rom{\'a}n-Garza}, {Tran}, \&
  {Zapartas}}]{bavera2021A&A...647A.153B}
{Bavera}, S.~S., {Fragos}, T., {Zevin}, M., {et~al.} 2021, \aap, 647, A153

\bibitem[{{Belczynski} {et~al.}(2002){Belczynski}, {Kalogera}, \&
  {Bulik}}]{Belczynski2002ApJ...572..407B}
{Belczynski}, K., {Kalogera}, V., \& {Bulik}, T. 2002, \apj, 572, 407

\bibitem[{{Belczynski} {et~al.}(2008){Belczynski}, {Kalogera}, {Rasio}, {Taam},
  {Zezas}, {Bulik}, {Maccarone}, \& {Ivanova}}]{Belczynski2008ApJS..174..223B}
{Belczynski}, K., {Kalogera}, V., {Rasio}, F.~A., {et~al.} 2008, \apjs, 174,
  223

\bibitem[{{Belczynski} {et~al.}(2020){Belczynski}, {Klencki}, {Fields},
  {Olejak}, {Berti}, {Meynet}, {Fryer}, {Holz}, {O'Shaughnessy}, {Brown},
  {Bulik}, {Leung}, {Nomoto}, {Madau}, {Hirschi}, {Kaiser}, {Jones}, {Mondal},
  {Chruslinska}, {Drozda}, {Gerosa}, {Doctor}, {Giersz}, {Ekstrom}, {Georgy},
  {Askar}, {Baibhav}, {Wysocki}, {Natan}, {Farr}, {Wiktorowicz}, {Coleman
  Miller}, {Farr}, \& {Lasota}}]{belczynski2020A&A...636A.104B}
{Belczynski}, K., {Klencki}, J., {Fields}, C.~E., {et~al.} 2020, \aap, 636,
  A104

\bibitem[{{Boesky} {et~al.}(2024){Boesky}, {Broekgaarden}, \&
  {Berger}}]{Boesky2024ApJ...976...24B}
{Boesky}, A.~P., {Broekgaarden}, F.~S., \& {Berger}, E. 2024, \apj, 976, 24

\bibitem[{{Breivik} {et~al.}(2020){Breivik}, {Coughlin}, {Zevin}, {Rodriguez},
  {Kremer}, {Ye}, {Andrews}, {Kurkowski}, {Digman}, {Larson}, \&
  {Rasio}}]{Breivik2020ApJ...898...71B}
{Breivik}, K., {Coughlin}, S., {Zevin}, M., {et~al.} 2020, \apj, 898, 71

\bibitem[{{Breivik} {et~al.}(2018){Breivik}, {Kremer}, {Bueno}, {Larson},
  {Coughlin}, \& {Kalogera}}]{Breivik2018ApJ...854L...1B}
{Breivik}, K., {Kremer}, K., {Bueno}, M., {et~al.} 2018, \apjl, 854, L1

\bibitem[{{Bressan} {et~al.}(2012){Bressan}, {Marigo}, {Girardi}, {Salasnich},
  {Dal Cero}, {Rubele}, \& {Nanni}}]{Bressan_2012}
{Bressan}, A., {Marigo}, P., {Girardi}, L., {et~al.} 2012, \mnras, 427, 127

\bibitem[{{Broekgaarden} {et~al.}(2021){Broekgaarden}, {Berger}, {Neijssel},
  {Vigna-G{\'o}mez}, {Chattopadhyay}, {Stevenson}, {Chruslinska}, {Justham},
  {de Mink}, \& {Mandel}}]{Broekgaarden2021MNRAS.508.5028B}
{Broekgaarden}, F.~S., {Berger}, E., {Neijssel}, C.~J., {et~al.} 2021, \mnras,
  508, 5028

\bibitem[{{Chatterjee} {et~al.}(2017){Chatterjee}, {Rodriguez}, \&
  {Rasio}}]{Chatterjee2017ApJ...834...68C}
{Chatterjee}, S., {Rodriguez}, C.~L., \& {Rasio}, F.~A. 2017, \apj, 834, 68

\bibitem[{{Chattopadhyay} {et~al.}(2021){Chattopadhyay}, {Stevenson}, {Hurley},
  {Bailes}, \& {Broekgaarden}}]{Chattopadhyay2021MNRAS.504.3682C}
{Chattopadhyay}, D., {Stevenson}, S., {Hurley}, J.~R., {Bailes}, M., \&
  {Broekgaarden}, F. 2021, \mnras, 504, 3682

\bibitem[{{Chen} \& {Han}(2008{\natexlab{a}})}]{Chen2008}
{Chen}, X. \& {Han}, Z. 2008{\natexlab{a}}, \mnras, 384, 1263

\bibitem[{{Chen} \& {Han}(2008{\natexlab{b}})}]{2008MNRAS.387.1416C}
{Chen}, X. \& {Han}, Z. 2008{\natexlab{b}}, \mnras, 387, 1416

\bibitem[{{Chen} {et~al.}(2015){Chen}, {Bressan}, {Girardi}, {Marigo}, {Kong},
  \& {Lanza}}]{Chen2015}
{Chen}, Y., {Bressan}, A., {Girardi}, L., {et~al.} 2015, \mnras, 452, 1068

\bibitem[{{Choi} {et~al.}(2016){Choi}, {Dotter}, {Conroy}, {Cantiello},
  {Paxton}, \& {Johnson}}]{choi2016ApJ...823..102C}
{Choi}, J., {Dotter}, A., {Conroy}, C., {et~al.} 2016, \apj, 823, 102

\bibitem[{{Claeys} {et~al.}(2014){Claeys}, {Pols}, {Izzard}, {Vink}, \&
  {Verbunt}}]{claeys2014A&A...563A..83C}
{Claeys}, J.~S.~W., {Pols}, O.~R., {Izzard}, R.~G., {Vink}, J., \& {Verbunt},
  F.~W.~M. 2014, \aap, 563, A83

\bibitem[{{Costa} {et~al.}(2023){Costa}, {Mapelli}, {Iorio}, {Santoliquido},
  {Escobar}, {Klessen}, \& {Bressan}}]{Costa2023}
{Costa}, G., {Mapelli}, M., {Iorio}, G., {et~al.} 2023, \mnras, 525, 2891

\bibitem[{{Costa} {et~al.}(2025){Costa}, {Shepherd}, {Bressan}, {Addari},
  {Chen}, {Fu}, {Volpato}, {Nguyen}, {Girardi}, {Marigo}, {Mazzi},
  {Pastorelli}, {Trabucchi}, {Bossini}, \& {Zaggia}}]{costa2025A&A...694A.193C}
{Costa}, G., {Shepherd}, K.~G., {Bressan}, A., {et~al.} 2025, \aap, 694, A193

\bibitem[{{De Donder} \& {Vanbeveren}(2004)}]{dedodner2004NewAR..48..861D}
{De Donder}, E. \& {Vanbeveren}, D. 2004, \nar, 48, 861

\bibitem[{{de Mink} {et~al.}(2007){de Mink}, {Pols}, \&
  {Hilditch}}]{DeMink2007}
{de Mink}, S.~E., {Pols}, O.~R., \& {Hilditch}, R.~W. 2007, \aap, 467, 1181

\bibitem[{{Dorozsmai} \& {Toonen}(2024)}]{dorozsmai2024MNRAS.530.3706D}
{Dorozsmai}, A. \& {Toonen}, S. 2024, \mnras, 530, 3706

\bibitem[{Echeveste \& Escobar(2026)}]{Echeveste2026}
Echeveste, M. \& Escobar, G.~J. 2026, AAA Workshop Series 14, in press

\bibitem[{{Eggleton}(1983)}]{Eggleton1983}
{Eggleton}, P.~P. 1983, \apj, 268, 368

\bibitem[{{Eldridge} {et~al.}(2008){Eldridge}, {Izzard}, \&
  {Tout}}]{Eldridge2008}
{Eldridge}, J.~J., {Izzard}, R.~G., \& {Tout}, C.~A. 2008, \mnras, 384, 1109

\bibitem[{{Eldridge} \& {Stanway}(2016)}]{2016MNRAS.462.3302E}
{Eldridge}, J.~J. \& {Stanway}, E.~R. 2016, \mnras, 462, 3302

\bibitem[{{Eldridge} {et~al.}(2017){Eldridge}, {Stanway}, {Xiao}, {McClelland},
  {Taylor}, {Ng}, {Greis}, \& {Bray}}]{Eldridge2017PASA...34...58E}
{Eldridge}, J.~J., {Stanway}, E.~R., {Xiao}, L., {et~al.} 2017, \pasa, 34, e058

\bibitem[{Escobar \& Echeveste(2026)}]{Escobar2026}
Escobar, G.~J. \& Echeveste, M. 2026, AAA Workshop Series 14, in press

\bibitem[{{Fragos} {et~al.}(2023){Fragos}, {Andrews}, {Bavera}, {Berry},
  {Coughlin}, {Dotter}, {Giri}, {Kalogera}, {Katsaggelos}, {Kovlakas},
  {Lalvani}, {Misra}, {Srivastava}, {Qin}, {Rocha}, {Rom{\'a}n-Garza}, {Serra},
  {Stahle}, {Sun}, {Teng}, {Trajcevski}, {Tran}, {Xing}, {Zapartas}, \&
  {Zevin}}]{Fragos2023}
{Fragos}, T., {Andrews}, J.~J., {Bavera}, S.~S., {et~al.} 2023, \apjs, 264, 45

\bibitem[{{Fryer} {et~al.}(2012){Fryer}, {Belczynski}, {Wiktorowicz},
  {Dominik}, {Kalogera}, \& {Holz}}]{Fryer2012ApJ...749...91F}
{Fryer}, C.~L., {Belczynski}, K., {Wiktorowicz}, G., {et~al.} 2012, \apj, 749,
  91

\bibitem[{{Ge} \& {Han}(2024)}]{Ge2024arXiv241117333G}
{Ge}, H. \& {Han}, Z. 2024, arXiv e-prints, arXiv:2411.17333

\bibitem[{{Ge} {et~al.}(2010){Ge}, {Hjellming}, {Webbink}, {Chen}, \&
  {Han}}]{ge2010ApJ...717..724G}
{Ge}, H., {Hjellming}, M.~S., {Webbink}, R.~F., {Chen}, X., \& {Han}, Z. 2010,
  \apj, 717, 724

\bibitem[{{Ge} {et~al.}(2022){Ge}, {Tout}, {Chen}, {Kruckow}, {Chen}, {Jiang},
  {Li}, {Liu}, \& {Han}}]{ge2022ApJ...933..137G}
{Ge}, H., {Tout}, C.~A., {Chen}, X., {et~al.} 2022, \apj, 933, 137

\bibitem[{{Ge} {et~al.}(2023){Ge}, {Tout}, {Chen}, {Sarkar}, {Walton}, \&
  {Han}}]{Ge2023ApJ...945....7G}
{Ge}, H., {Tout}, C.~A., {Chen}, X., {et~al.} 2023, \apj, 945, 7

\bibitem[{{Ge} {et~al.}(2024{\natexlab{a}}){Ge}, {Tout}, {Chen}, {Wang},
  {Xiong}, {Zhang}, {Li}, {Liu}, \& {Han}}]{Ge2024ApJ...975..254G}
{Ge}, H., {Tout}, C.~A., {Chen}, X., {et~al.} 2024{\natexlab{a}}, \apj, 975,
  254

\bibitem[{{Ge} {et~al.}(2024{\natexlab{b}}){Ge}, {Tout}, {Webbink}, {Chen},
  {Sarkar}, {Li}, {Li}, {Zhang}, \& {Han}}]{ge2024ApJ...961..202G}
{Ge}, H., {Tout}, C.~A., {Webbink}, R.~F., {et~al.} 2024{\natexlab{b}}, \apj,
  961, 202

\bibitem[{{Ge} {et~al.}(2015){Ge}, {Webbink}, {Chen}, \&
  {Han}}]{Ge2015ApJ...812...40G}
{Ge}, H., {Webbink}, R.~F., {Chen}, X., \& {Han}, Z. 2015, \apj, 812, 40

\bibitem[{{Ge} {et~al.}(2020){Ge}, {Webbink}, {Chen}, \&
  {Han}}]{Ge2020ApJ...899..132G}
{Ge}, H., {Webbink}, R.~F., {Chen}, X., \& {Han}, Z. 2020, \apj, 899, 132

\bibitem[{{Geller} {et~al.}(2015){Geller}, {Latham}, \& {Mathieu}}]{geller2015}
{Geller}, A.~M., {Latham}, D.~W., \& {Mathieu}, R.~D. 2015, \aj, 150, 97

\bibitem[{{Geller} \& {Mathieu}(2011)}]{Geller_2011}
{Geller}, A.~M. \& {Mathieu}, R.~D. 2011, \nat, 478, 356

\bibitem[{{Geller} \& {Mathieu}(2012)}]{Geller_2012}
{Geller}, A.~M. \& {Mathieu}, R.~D. 2012, \aj, 144, 54

\bibitem[{{Giacobbo} \& {Mapelli}(2018)}]{Giacobbo2018}
{Giacobbo}, N. \& {Mapelli}, M. 2018, \mnras, 480, 2011

\bibitem[{{Giacobbo} \& {Mapelli}(2020)}]{Giacobbo2020ApJ...891..141G}
{Giacobbo}, N. \& {Mapelli}, M. 2020, \apj, 891, 141

\bibitem[{{Giacobbo} {et~al.}(2018){Giacobbo}, {Mapelli}, \&
  {Spera}}]{giacobbo2018MNRAS.474.2959G}
{Giacobbo}, N., {Mapelli}, M., \& {Spera}, M. 2018, \mnras, 474, 2959

\bibitem[{{Gosnell} {et~al.}(2015){Gosnell}, {Mathieu}, {Geller}, {Sills},
  {Leigh}, \& {Knigge}}]{Gosnell_2015}
{Gosnell}, N.~M., {Mathieu}, R.~D., {Geller}, A.~M., {et~al.} 2015, \apj, 814,
  163

\bibitem[{{Hjellming} \& {Webbink}(1987)}]{Hjellming1987}
{Hjellming}, M.~S. \& {Webbink}, R.~F. 1987, \apj, 318, 794

\bibitem[{{Hurley} {et~al.}(2005){Hurley}, {Pols}, {Aarseth}, \&
  {Tout}}]{2005MNRAS.363..293H}
{Hurley}, J.~R., {Pols}, O.~R., {Aarseth}, S.~J., \& {Tout}, C.~A. 2005,
  \mnras, 363, 293

\bibitem[{{Hurley} {et~al.}(2000){Hurley}, {Pols}, \& {Tout}}]{Hurley2000}
{Hurley}, J.~R., {Pols}, O.~R., \& {Tout}, C.~A. 2000, \mnras, 315, 543

\bibitem[{{Hurley} {et~al.}(2002){Hurley}, {Tout}, \& {Pols}}]{Hurley2002}
{Hurley}, J.~R., {Tout}, C.~A., \& {Pols}, O.~R. 2002, \mnras, 329, 897

\bibitem[{{Hut}(1981)}]{Hut1981A&A....99..126H}
{Hut}, P. 1981, \aap, 99, 126

\bibitem[{{Iorio} {et~al.}(2023){Iorio}, {Mapelli}, {Costa}, {Spera},
  {Escobar}, {Sgalletta}, {Trani}, {Korb}, {Santoliquido}, {Dall'Amico},
  {Gaspari}, \& {Bressan}}]{Iorio2023}
{Iorio}, G., {Mapelli}, M., {Costa}, G., {et~al.} 2023, \mnras, 524, 426

\bibitem[{{Ivanova} {et~al.}(2013){Ivanova}, {Justham}, {Chen}, {De Marco},
  {Fryer}, {Gaburov}, {Ge}, {Glebbeek}, {Han}, {Li}, {Lu}, {Marsh},
  {Podsiadlowski}, {Potter}, {Soker}, {Taam}, {Tauris}, {van den Heuvel}, \&
  {Webbink}}]{Ivanova2013A&ARv..21...59I}
{Ivanova}, N., {Justham}, S., {Chen}, X., {et~al.} 2013, \aapr, 21, 59

\bibitem[{{Izzard} {et~al.}(2006){Izzard}, {Dray}, {Karakas}, {Lugaro}, \&
  {Tout}}]{izzard2006A&A...460..565I}
{Izzard}, R.~G., {Dray}, L.~M., {Karakas}, A.~I., {Lugaro}, M., \& {Tout},
  C.~A. 2006, \aap, 460, 565

\bibitem[{{Izzard} {et~al.}(2009){Izzard}, {Glebbeek}, {Stancliffe}, \&
  {Pols}}]{Izzard2009}
{Izzard}, R.~G., {Glebbeek}, E., {Stancliffe}, R.~J., \& {Pols}, O.~R. 2009,
  \aap, 508, 1359

\bibitem[{{Izzard} {et~al.}(2018){Izzard}, {Preece}, {Jofre}, {Halabi},
  {Masseron}, \& {Tout}}]{Izzard2018}
{Izzard}, R.~G., {Preece}, H., {Jofre}, P., {et~al.} 2018, \mnras, 473, 2984

\bibitem[{{Izzard} {et~al.}(2004){Izzard}, {Tout}, {Karakas}, \&
  {Pols}}]{Izzard2004}
{Izzard}, R.~G., {Tout}, C.~A., {Karakas}, A.~I., \& {Pols}, O.~R. 2004,
  \mnras, 350, 407

\bibitem[{{Klencki} {et~al.}(2021){Klencki}, {Nelemans}, {Istrate}, \&
  {Chruslinska}}]{Klencki2021A&A...645A..54K}
{Klencki}, J., {Nelemans}, G., {Istrate}, A.~G., \& {Chruslinska}, M. 2021,
  \aap, 645, A54

\bibitem[{{Klencki} {et~al.}(2020){Klencki}, {Nelemans}, {Istrate}, \&
  {Pols}}]{Klencki2020A&A...638A..55K}
{Klencki}, J., {Nelemans}, G., {Istrate}, A.~G., \& {Pols}, O. 2020, \aap, 638,
  A55

\bibitem[{{Knigge} {et~al.}(2009){Knigge}, {Leigh}, \& {Sills}}]{Knigge_2009}
{Knigge}, C., {Leigh}, N., \& {Sills}, A. 2009, \nat, 457, 288

\bibitem[{{Korol} {et~al.}(2022){Korol}, {Belokurov}, \&
  {Toonen}}]{korol2022MNRAS.515.1228K}
{Korol}, V., {Belokurov}, V., \& {Toonen}, S. 2022, \mnras, 515, 1228

\bibitem[{{Korol} {et~al.}(2020){Korol}, {Toonen}, {Klein}, {Belokurov},
  {Vincenzo}, {Buscicchio}, {Gerosa}, {Moore}, {Roebber}, {Rossi}, \&
  {Vecchio}}]{korol2020A&A...638A.153K}
{Korol}, V., {Toonen}, S., {Klein}, A., {et~al.} 2020, \aap, 638, A153

\bibitem[{{Kremer} {et~al.}(2020){Kremer}, {Ye}, {Rui}, {Weatherford},
  {Chatterjee}, {Fragione}, {Rodriguez}, {Spera}, \&
  {Rasio}}]{kremer2020ApJS..247...48K}
{Kremer}, K., {Ye}, C.~S., {Rui}, N.~Z., {et~al.} 2020, \apjs, 247, 48

\bibitem[{{Kroupa}(2001)}]{Kroupa2001}
{Kroupa}, P. 2001, \mnras, 322, 231

\bibitem[{{Kruckow} {et~al.}(2018){Kruckow}, {Tauris}, {Langer}, {Kramer}, \&
  {Izzard}}]{Kruckow2018}
{Kruckow}, M.~U., {Tauris}, T.~M., {Langer}, N., {Kramer}, M., \& {Izzard},
  R.~G. 2018, \mnras, 481, 1908

\bibitem[{{Lecroq} {et~al.}(2024){Lecroq}, {Charlot}, {Bressan}, {Bruzual},
  {Costa}, {Iorio}, {Spera}, {Mapelli}, {Chen}, {Chevallard}, \&
  {Dall'Amico}}]{Lecroq2024MNRAS.527.9480L}
{Lecroq}, M., {Charlot}, S., {Bressan}, A., {et~al.} 2024, \mnras, 527, 9480

\bibitem[{{Leiner} \& {Geller}(2021)}]{Leiner_2021}
{Leiner}, E.~M. \& {Geller}, A. 2021, \apj, 908, 229

\bibitem[{{Li} {et~al.}(2025){Li}, {L{\"u}}, {Zhu}, {Guo}, {Ge}, {Gu}, {Li}, \&
  {He}}]{li2025PhRvD.112j3005L}
{Li}, L., {L{\"u}}, G., {Zhu}, C., {et~al.} 2025, \prd, 112, 103005

\bibitem[{{Li} {et~al.}(2023){Li}, {Chen}, {Ge}, {Chen}, \&
  {Han}}]{li2023A&A...669A..82L}
{Li}, Z., {Chen}, X., {Ge}, H., {Chen}, H.-L., \& {Han}, Z. 2023, \aap, 669,
  A82

\bibitem[{{Loveridge} {et~al.}(2011){Loveridge}, {van der Sluys}, \&
  {Kalogera}}]{Loveridge2011ApJ...743...49L}
{Loveridge}, A.~J., {van der Sluys}, M.~V., \& {Kalogera}, V. 2011, \apj, 743,
  49

\bibitem[{{Mapelli} {et~al.}(2017){Mapelli}, {Giacobbo}, {Ripamonti}, \&
  {Spera}}]{Mapelli2017}
{Mapelli}, M., {Giacobbo}, N., {Ripamonti}, E., \& {Spera}, M. 2017, \mnras,
  472, 2422

\bibitem[{{Mapelli} {et~al.}(2020){Mapelli}, {Spera}, {Montanari}, {Limongi},
  {Chieffi}, {Giacobbo}, {Bressan}, \& {Bouffanais}}]{Mapelli2020}
{Mapelli}, M., {Spera}, M., {Montanari}, E., {et~al.} 2020, \apj, 888, 76

\bibitem[{{Marchant} {et~al.}(2021){Marchant}, {Pappas}, {Gallegos-Garcia},
  {Berry}, {Taam}, {Kalogera}, \& {Podsiadlowski}}]{Marchant2021}
{Marchant}, P., {Pappas}, K. M.~W., {Gallegos-Garcia}, M., {et~al.} 2021, \aap,
  650, A107

\bibitem[{{Mar{\'\i}n Pina} {et~al.}(2024){Mar{\'\i}n Pina}, {Rastello},
  {Gieles}, {Kremer}, {Fitzgerald}, \& {Rando Forastier}}]{MarinPina2024}
{Mar{\'\i}n Pina}, D., {Rastello}, S., {Gieles}, M., {et~al.} 2024, \aap, 688,
  L2

\bibitem[{{Marinacci} {et~al.}(2025){Marinacci}, {Baldi}, {Iorio}, {Artale},
  {Mapelli}, {Springel}, {Bose}, \& {Hernquist}}]{Marinacci2025}
{Marinacci}, F., {Baldi}, M., {Iorio}, G., {et~al.} 2025, arXiv e-prints,
  arXiv:2510.06311

\bibitem[{{Mathieu} \& {Geller}(2009)}]{Mathieu_2009}
{Mathieu}, R.~D. \& {Geller}, A.~M. 2009, \nat, 462, 1032

\bibitem[{{McCrea}(1964)}]{McCrea_1964}
{McCrea}, W.~H. 1964, \mnras, 128, 147

\bibitem[{{Mestichelli} {et~al.}(2024){Mestichelli}, {Mapelli}, {Torniamenti},
  {Arca Sedda}, {Branchesi}, {Costa}, {Iorio}, \&
  {Santoliquido}}]{Mestichelli2024}
{Mestichelli}, B., {Mapelli}, M., {Torniamenti}, S., {et~al.} 2024, \aap, 690,
  A106

\bibitem[{{Neijssel} {et~al.}(2019){Neijssel}, {Vigna-G{\'o}mez}, {Stevenson},
  {Barrett}, {Gaebel}, {Broekgaarden}, {de Mink}, {Sz{\'e}csi}, {Vinciguerra},
  \& {Mandel}}]{Neijssel2019MNRAS.490.3740N}
{Neijssel}, C.~J., {Vigna-G{\'o}mez}, A., {Stevenson}, S., {et~al.} 2019,
  \mnras, 490, 3740

\bibitem[{{Nelemans} {et~al.}(2001){Nelemans}, {Yungelson}, {Portegies Zwart},
  \& {Verbunt}}]{Nelemans2001A&A...365..491N}
{Nelemans}, G., {Yungelson}, L.~R., {Portegies Zwart}, S.~F., \& {Verbunt}, F.
  2001, \aap, 365, 491

\bibitem[{{Nie} {et~al.}(2012){Nie}, {Wood}, \&
  {Nicholls}}]{nie2012MNRAS.423.2764N}
{Nie}, J.~D., {Wood}, P.~R., \& {Nicholls}, C.~P. 2012, \mnras, 423, 2764

\bibitem[{Nitz {et~al.}(2020)Nitz, Dent, Davies, Kumar, Capano, Harry, Mozzon,
  Nuttall, Lundgren, \& Tápai}]{Nitz_2020}
Nitz, A.~H., Dent, T., Davies, G.~S., {et~al.} 2020, The Astrophysical Journal,
  891, 123

\bibitem[{{Olejak} {et~al.}(2021){Olejak}, {Belczynski}, \&
  {Ivanova}}]{Olejak2021}
{Olejak}, A., {Belczynski}, K., \& {Ivanova}, N. 2021, \aap, 651, A100

\bibitem[{{Paczynski}(1976)}]{Paczynski1976IAUS...73...75P}
{Paczynski}, B. 1976, in IAU Symposium, Vol.~73, Structure and Evolution of
  Close Binary Systems, ed. P.~{Eggleton}, S.~{Mitton}, \& J.~{Whelan}, 75

\bibitem[{{Pandey} {et~al.}(2021){Pandey}, {Subramaniam}, \&
  {Jadhav}}]{pandey2021}
{Pandey}, S., {Subramaniam}, A., \& {Jadhav}, V.~V. 2021, \mnras, 507, 2373

\bibitem[{{Pavlovskii} {et~al.}(2017){Pavlovskii}, {Ivanova}, {Belczynski}, \&
  {Van}}]{Pavlovskii2017MNRAS.465.2092P}
{Pavlovskii}, K., {Ivanova}, N., {Belczynski}, K., \& {Van}, K.~X. 2017,
  \mnras, 465, 2092

\bibitem[{{Paxton} {et~al.}(2011){Paxton}, {Bildsten}, {Dotter}, {Herwig},
  {Lesaffre}, \& {Timmes}}]{Paxton2011}
{Paxton}, B., {Bildsten}, L., {Dotter}, A., {et~al.} 2011, \apjs, 192, 3

\bibitem[{{Paxton} {et~al.}(2013){Paxton}, {Cantiello}, {Arras}, {Bildsten},
  {Brown}, {Dotter}, {Mankovich}, {Montgomery}, {Stello}, {Timmes}, \&
  {Townsend}}]{Paxton2013}
{Paxton}, B., {Cantiello}, M., {Arras}, P., {et~al.} 2013, \apjs, 208, 4

\bibitem[{{Paxton} {et~al.}(2015){Paxton}, {Marchant}, {Schwab}, {Bauer},
  {Bildsten}, {Cantiello}, {Dessart}, {Farmer}, {Hu}, {Langer}, {Townsend},
  {Townsley}, \& {Timmes}}]{Paxton2015}
{Paxton}, B., {Marchant}, P., {Schwab}, J., {et~al.} 2015, \apjs, 220, 15

\bibitem[{{Peters}(1964)}]{Peters1964}
{Peters}, P.~C. 1964, Physical Review, 136, 1224

\bibitem[{{Picco} {et~al.}(2024){Picco}, {Marchant}, {Sana}, \&
  {Nelemans}}]{picco2024A&A...681A..31P}
{Picco}, A., {Marchant}, P., {Sana}, H., \& {Nelemans}, G. 2024, \aap, 681, A31

\bibitem[{{Planck Collaboration} {et~al.}(2020){Planck Collaboration},
  {Aghanim}, {Akrami}, {Ashdown}, {Aumont}, {Baccigalupi}, {Ballardini},
  {Banday}, {Barreiro}, {Bartolo}, {Basak}, {Battye}, {Benabed}, {Bernard},
  {Bersanelli}, {Bielewicz}, {Bock}, {Bond}, {Borrill}, {Bouchet}, {Boulanger},
  {Bucher}, {Burigana}, {Butler}, {Calabrese}, {Cardoso}, {Carron},
  {Challinor}, {Chiang}, {Chluba}, {Colombo}, {Combet}, {Contreras}, {Crill},
  {Cuttaia}, {de Bernardis}, {de Zotti}, {Delabrouille}, {Delouis}, {Di
  Valentino}, {Diego}, {Dor{\'e}}, {Douspis}, {Ducout}, {Dupac}, {Dusini},
  {Efstathiou}, {Elsner}, {En{\ss}lin}, {Eriksen}, {Fantaye}, {Farhang},
  {Fergusson}, {Fernandez-Cobos}, {Finelli}, {Forastieri}, {Frailis},
  {Fraisse}, {Franceschi}, {Frolov}, {Galeotta}, {Galli}, {Ganga},
  {G{\'e}nova-Santos}, {Gerbino}, {Ghosh}, {Gonz{\'a}lez-Nuevo}, {G{\'o}rski},
  {Gratton}, {Gruppuso}, {Gudmundsson}, {Hamann}, {Handley}, {Hansen},
  {Herranz}, {Hildebrandt}, {Hivon}, {Huang}, {Jaffe}, {Jones}, {Karakci},
  {Keih{\"a}nen}, {Keskitalo}, {Kiiveri}, {Kim}, {Kisner}, {Knox},
  {Krachmalnicoff}, {Kunz}, {Kurki-Suonio}, {Lagache}, {Lamarre}, {Lasenby},
  {Lattanzi}, {Lawrence}, {Le Jeune}, {Lemos}, {Lesgourgues}, {Levrier},
  {Lewis}, {Liguori}, {Lilje}, {Lilley}, {Lindholm}, {L{\'o}pez-Caniego},
  {Lubin}, {Ma}, {Mac{\'\i}as-P{\'e}rez}, {Maggio}, {Maino}, {Mandolesi},
  {Mangilli}, {Marcos-Caballero}, {Maris}, {Martin}, {Martinelli},
  {Mart{\'\i}nez-Gonz{\'a}lez}, {Matarrese}, {Mauri}, {McEwen}, {Meinhold},
  {Melchiorri}, {Mennella}, {Migliaccio}, {Millea}, {Mitra},
  {Miville-Desch{\^e}nes}, {Molinari}, {Montier}, {Morgante}, {Moss}, {Natoli},
  {N{\o}rgaard-Nielsen}, {Pagano}, {Paoletti}, {Partridge}, {Patanchon},
  {Peiris}, {Perrotta}, {Pettorino}, {Piacentini}, {Polastri}, {Polenta},
  {Puget}, {Rachen}, {Reinecke}, {Remazeilles}, {Renzi}, {Rocha}, {Rosset},
  {Roudier}, {Rubi{\~n}o-Mart{\'\i}n}, {Ruiz-Granados}, {Salvati}, {Sandri},
  {Savelainen}, {Scott}, {Shellard}, {Sirignano}, {Sirri}, {Spencer},
  {Sunyaev}, {Suur-Uski}, {Tauber}, {Tavagnacco}, {Tenti}, {Toffolatti},
  {Tomasi}, {Trombetti}, {Valenziano}, {Valiviita}, {Van Tent}, {Vibert},
  {Vielva}, {Villa}, {Vittorio}, {Wandelt}, {Wehus}, {White}, {White},
  {Zacchei}, \& {Zonca}}]{Planck2020}
{Planck Collaboration}, {Aghanim}, N., {Akrami}, Y., {et~al.} 2020, \aap, 641,
  A6

\bibitem[{{Pols}(2018)}]{Pols2018}
{Pols}, O. 2018, 'Course Notes-Binary Stars',
  \\url{https://www.astro.ru.nl/~onnop/}

\bibitem[{{Pols} {et~al.}(1998){Pols}, {Schr{\"o}der}, {Hurley}, {Tout}, \&
  {Eggleton}}]{Pols1998}
{Pols}, O.~R., {Schr{\"o}der}, K.-P., {Hurley}, J.~R., {Tout}, C.~A., \&
  {Eggleton}, P.~P. 1998, \mnras, 298, 525

\bibitem[{{Popham} \& {Narayan}(1991)}]{Popham1991}
{Popham}, R. \& {Narayan}, R. 1991, \apj, 370, 604

\bibitem[{{Portegies Zwart} \& {Verbunt}(1996)}]{Portegies1996A&A...309..179P}
{Portegies Zwart}, S.~F. \& {Verbunt}, F. 1996, \aap, 309, 179

\bibitem[{{Rain} {et~al.}(2024){Rain}, {Pera}, {Perren}, {Benvenuto}, {Panei},
  {De Vito}, {Carraro}, \& {Villanova}}]{Rain_2024}
{Rain}, M.~J., {Pera}, M.~S., {Perren}, G.~I., {et~al.} 2024, \aap, 685, A33

\bibitem[{{Rastello} {et~al.}(2023){Rastello}, {Iorio}, {Mapelli},
  {Arca-Sedda}, {Di Carlo}, {Escobar}, {Shenar}, \&
  {Torniamenti}}]{Rastello2023}
{Rastello}, S., {Iorio}, G., {Mapelli}, M., {et~al.} 2023, \mnras, 526, 740

\bibitem[{{Riley} {et~al.}(2022){Riley}, {Agrawal}, {Barrett}, {Boyett},
  {Broekgaarden}, {Chattopadhyay}, {Gaebel}, {Gittins}, {Hirai}, {Howitt},
  {Justham}, {Khandelwal}, {Kummer}, {Lau}, {Mandel}, {de Mink}, {Neijssel},
  {Riley}, {van Son}, {Stevenson}, {Vigna-G{\'o}mez}, {Vinciguerra}, {Wagg},
  {Willcox}, \& {Team Compas}}]{Riley2022}
{Riley}, J., {Agrawal}, P., {Barrett}, J.~W., {et~al.} 2022, \apjs, 258, 34

\bibitem[{{Riley} {et~al.}(2021){Riley}, {Mandel}, {Marchant}, {Butler},
  {Nathaniel}, {Neijssel}, {Shortt}, \& {Vigna-G{\'o}mez}}]{Riley2021}
{Riley}, J., {Mandel}, I., {Marchant}, P., {et~al.} 2021, \mnras, 505, 663

\bibitem[{{Rodriguez} {et~al.}(2016){Rodriguez}, {Chatterjee}, \&
  {Rasio}}]{rodriguez2016PhRvD..93h4029R}
{Rodriguez}, C.~L., {Chatterjee}, S., \& {Rasio}, F.~A. 2016, \prd, 93, 084029

\bibitem[{{Rodr{\'\i}guez-Segovia} \& {Ruiter}(2025)}]{Rodriguez-Segovia2025_I}
{Rodr{\'\i}guez-Segovia}, N. \& {Ruiter}, A.~J. 2025, \mnras, 539, 3273

\bibitem[{{Rodr{\'\i}guez-Segovia} {et~al.}(2025){Rodr{\'\i}guez-Segovia},
  {Ruiter}, \& {Seitenzahl}}]{Rodriguez-Segovia2025_II}
{Rodr{\'\i}guez-Segovia}, N., {Ruiter}, A.~J., \& {Seitenzahl}, I.~R. 2025,
  \pasa, 42, e012

\bibitem[{{Rom{\'a}n-Garza} {et~al.}(2021){Rom{\'a}n-Garza}, {Bavera},
  {Fragos}, {Zapartas}, {Misra}, {Andrews}, {Coughlin}, {Dotter}, {Kovlakas},
  {Serra}, {Qin}, {Rocha}, \& {Tran}}]{ramon2021ApJ...912L..23R}
{Rom{\'a}n-Garza}, J., {Bavera}, S.~S., {Fragos}, T., {et~al.} 2021, \apjl,
  912, L23

\bibitem[{{Sana} {et~al.}(2012){Sana}, {de Mink}, {de Koter}, {Langer},
  {Evans}, {Gieles}, {Gosset}, {Izzard}, {Le Bouquin}, \&
  {Schneider}}]{Sana2012}
{Sana}, H., {de Mink}, S.~E., {de Koter}, A., {et~al.} 2012, Science, 337, 444

\bibitem[{{Sandage}(1953)}]{sandage53}
{Sandage}, A. 1953, PhD thesis, California Institute of Technology

\bibitem[{{Scherbak} \& {Fuller}(2023)}]{Scherbak2023MNRAS.518.3966S}
{Scherbak}, P. \& {Fuller}, J. 2023, \mnras, 518, 3966

\bibitem[{{Sgalletta} {et~al.}(2025){Sgalletta}, {Mapelli}, {Boco},
  {Santoliquido}, {Artale}, {Iorio}, {Lapi}, \& {Spera}}]{Sgalletta2025}
{Sgalletta}, C., {Mapelli}, M., {Boco}, L., {et~al.} 2025, \aap, 698, A144

\bibitem[{{Shao} \& {Li}(2014)}]{Shao2014}
{Shao}, Y. \& {Li}, X.-D. 2014, \apj, 796, 37

\bibitem[{{Soberman} {et~al.}(1997){Soberman}, {Phinney}, \& {van den
  Heuvel}}]{soberman1997}
{Soberman}, G.~E., {Phinney}, E.~S., \& {van den Heuvel}, E.~P.~J. 1997, \aap,
  327, 620

\bibitem[{{Sollima} {et~al.}(2008){Sollima}, {Lanzoni}, {Beccari}, {Ferraro},
  \& {Fusi Pecci}}]{Sollima_2008}
{Sollima}, A., {Lanzoni}, B., {Beccari}, G., {Ferraro}, F.~R., \& {Fusi Pecci},
  F. 2008, \aap, 481, 701

\bibitem[{{Spera} \& {Mapelli}(2017)}]{spera2017MNRAS.470.4739S}
{Spera}, M. \& {Mapelli}, M. 2017, \mnras, 470, 4739

\bibitem[{{Spera} {et~al.}(2019){Spera}, {Mapelli}, {Giacobbo}, {Trani},
  {Bressan}, \& {Costa}}]{Spera2019}
{Spera}, M., {Mapelli}, M., {Giacobbo}, N., {et~al.} 2019, \mnras, 485, 889

\bibitem[{{Stevenson} {et~al.}(2017){Stevenson}, {Vigna-G{\'o}mez}, {Mandel},
  {Barrett}, {Neijssel}, {Perkins}, \& {de
  Mink}}]{Stevenson2017NatCo...814906S}
{Stevenson}, S., {Vigna-G{\'o}mez}, A., {Mandel}, I., {et~al.} 2017, Nature
  Communications, 8, 14906

\bibitem[{{Tauris} \& {van den Heuvel}(2023)}]{2023pbse.book.....T}
{Tauris}, T.~M. \& {van den Heuvel}, E. P.~J. 2023, {Physics of Binary Star
  Evolution. From Stars to X-ray Binaries and Gravitational Wave Sources}

\bibitem[{{Tian} {et~al.}(2006){Tian}, {Deng}, {Han}, \& {Zhang}}]{Tian_2006}
{Tian}, B., {Deng}, L., {Han}, Z., \& {Zhang}, X.~B. 2006, \aap, 455, 247

\bibitem[{{Toonen} {et~al.}(2012){Toonen}, {Nelemans}, \& {Portegies
  Zwart}}]{Toonen2012A&A...546A..70T}
{Toonen}, S., {Nelemans}, G., \& {Portegies Zwart}, S. 2012, \aap, 546, A70

\bibitem[{{Torniamenti} {et~al.}(2022){Torniamenti}, {Rastello}, {Mapelli}, {Di
  Carlo}, {Ballone}, \& {Pasquato}}]{Torniamenti2022}
{Torniamenti}, S., {Rastello}, S., {Mapelli}, M., {et~al.} 2022, \mnras, 517,
  2953

\bibitem[{{Tout} {et~al.}(1997){Tout}, {Aarseth}, {Pols}, \&
  {Eggleton}}]{Tout1997}
{Tout}, C.~A., {Aarseth}, S.~J., {Pols}, O.~R., \& {Eggleton}, P.~P. 1997,
  \mnras, 291, 732

\bibitem[{{Vaidya} {et~al.}(2022){Vaidya}, {Panthi}, {Agarwal}, {Pandey},
  {Rao}, {Jadhav}, \& {Subramaniam}}]{Vaidya_2022b}
{Vaidya}, K., {Panthi}, A., {Agarwal}, M., {et~al.} 2022, \mnras, 511, 2274

\bibitem[{{van Son} {et~al.}(2022){van Son}, {de Mink}, {Callister}, {Justham},
  {Renzo}, {Wagg}, {Broekgaarden}, {Kummer}, {Pakmor}, \&
  {Mandel}}]{vanSon2022ApJ...931...17V}
{van Son}, L.~A.~C., {de Mink}, S.~E., {Callister}, T., {et~al.} 2022, \apj,
  931, 17

\bibitem[{Venumadhav {et~al.}(2020)Venumadhav, Zackay, Roulet, Dai, \&
  Zaldarriaga}]{Venumadhav2020PhysRevD.101.083030}
Venumadhav, T., Zackay, B., Roulet, J., Dai, L., \& Zaldarriaga, M. 2020, Phys.
  Rev. D, 101, 083030

\bibitem[{{Vigna-G{\'o}mez} {et~al.}(2018){Vigna-G{\'o}mez}, {Neijssel},
  {Stevenson}, {Barrett}, {Belczynski}, {Justham}, {de Mink}, {M{\"u}ller},
  {Podsiadlowski}, {Renzo}, {Sz{\'e}csi}, \&
  {Mandel}}]{VignaGomez2018MNRAS.481.4009V}
{Vigna-G{\'o}mez}, A., {Neijssel}, C.~J., {Stevenson}, S., {et~al.} 2018,
  \mnras, 481, 4009

\bibitem[{{Wang} {et~al.}(2022){Wang}, {Langer}, {Schootemeijer}, {Milone},
  {Hastings}, {Xu}, {Bodensteiner}, {Sana}, {Castro}, {Lennon}, {Marchant}, {de
  Koter}, \& {de Mink}}]{Wang2022NatAs...6..480W}
{Wang}, C., {Langer}, N., {Schootemeijer}, A., {et~al.} 2022, Nature Astronomy,
  6, 480

\bibitem[{{Webbink}(1985)}]{webbink1985ibs..book...39W}
{Webbink}, R.~F. 1985, in Interacting Binary Stars, ed. J.~E. {Pringle} \&
  R.~A. {Wade}, 39

\bibitem[{Webbink(1988)}]{Webbink1988}
Webbink, R.~F. 1988, in The Symbiotic Phenomenon, ed. J.~Mikolajewska,
  M.~Friedjung, S.~J. Kenyon, \& R.~Viotti (Dordrecht: Springer Netherlands),
  311--321

\bibitem[{{Willcox} {et~al.}(2023){Willcox}, {MacLeod}, {Mandel}, \&
  {Hirai}}]{Willcox2023ApJ...958..138W}
{Willcox}, R., {MacLeod}, M., {Mandel}, I., \& {Hirai}, R. 2023, \apj, 958, 138

\bibitem[{{Xu} \& {Li}(2010)}]{xu2010ApJ...716..114X}
{Xu}, X.-J. \& {Li}, X.-D. 2010, \apj, 716, 114

\bibitem[{Zackay {et~al.}(2019)Zackay, Venumadhav, Dai, Roulet, \&
  Zaldarriaga}]{Zackay2019PhysRevD.100.023007}
Zackay, B., Venumadhav, T., Dai, L., Roulet, J., \& Zaldarriaga, M. 2019, Phys.
  Rev. D, 100, 023007

\bibitem[{{Zorotovic} \& {Schreiber}(2022)}]{Zorotovic2022MNRAS.513.3587Z}
{Zorotovic}, M. \& {Schreiber}, M. 2022, \mnras, 513, 3587

\end{thebibliography}

\end{document}